\documentclass[journal]{IEEEtran}

\usepackage{makecell}
\usepackage{float}
\usepackage{caption}
\usepackage{subcaption}
\usepackage{longtable}
\usepackage{tabularx}
\usepackage{tabu}
\usepackage{multirow}
\usepackage{tikz}
\usepackage{graphicx}
\usepackage{tikzscale}
\usepackage{pgfplots}
\usepackage{setspace}
\usepackage{booktabs}
\usepackage{siunitx}
\usepackage{adjustbox}
\usepackage{soul, color}
\usepackage{xcolor}
\usepackage[labelformat=simple]{subcaption}

\usepackage[most]{tcolorbox}
\newtcolorbox{highlighted}{colback=yellow,coltext=red,breakable}

\usepackage{algorithmic}
\usepackage{amsmath,amssymb,amsfonts}
\usepackage[ruled,linesnumbered]{algorithm2e}
\usepackage[font=small,labelfont=bf]{caption}
\SetKwFor{Foreach}{for each}{do}{endForEach}%
\SetKwFor{For}{for}{do}{endfor for}%
\SetKwIF{uIf}{ElseIf}{Else}{if}{then}{else if}{else}{endIf if}%
\SetKwFor{While}{while}{do}{endWhile while}%

\begin{document}

\title{
Blockchain-based Federated Learning with SMPC Model Verification Against Poisoning Attack for Healthcare Systems
}

\author{
Aditya Pribadi Kalapaaking, Ibrahim Khalil, Xun Yi
}

\maketitle

\begin{abstract}
Due to the rising awareness of privacy and security in machine learning applications, federated learning (FL) has received widespread attention and applied to several areas, e.g., intelligence healthcare systems, IoT-based industries, and smart cities. FL enables clients to train a global model collaboratively without accessing their local training data. However, the current FL schemes are vulnerable to adversarial attacks. Its architecture makes detecting and defending against malicious model updates difficult. In addition, most recent studies to detect FL from malicious updates while maintaining the model's privacy have not been sufficiently explored. This paper proposed blockchain-based federated learning with SMPC model verification against poisoning attacks for healthcare systems. First, we check the machine learning model from the FL participants through an encrypted inference process and remove the compromised model. Once the participants' local models have been verified, the models are sent to the blockchain node to be securely aggregated. We conducted several experiments with different medical datasets to evaluate our proposed framework.
\end{abstract}


% \fi
\IEEEpeerreviewmaketitle

\section{Introduction}

The Internet of Things (IoT) has been applied in various services, including the healthcare domain. The integration of IoT in the healthcare system is also known as the Internet of Medical Things (IoMT). With the development of IoMT, many healthcare devices are interconnected, allowing devices to exchange information among medical experts and Artificial Intelligence (AI) based services. This interconnectivity helps healthcare industries like hospitals to improve the efficiency and quality of their services. In the medical diagnosis field, medical imaging devices facilitate the process of early diagnosis and treatment for medical staff.

Due to this interconnectivity, medical image retrieval is made easy, resulting in extensive data with wide variations. Consequently, medical image analysis has become a challenging task for medical experts and is prone to human error. In recent years, the success of Deep Learning (DL) in computer vision tasks has provided a significant breakthrough in medical image classification tasks. Several studies of DL in medical imaging fields have shown promising results by providing accurate and efficient diagnoses \cite{sun2020edge}. 

\begin{figure}[tbh!]
\centering
\includegraphics[width=1\linewidth]{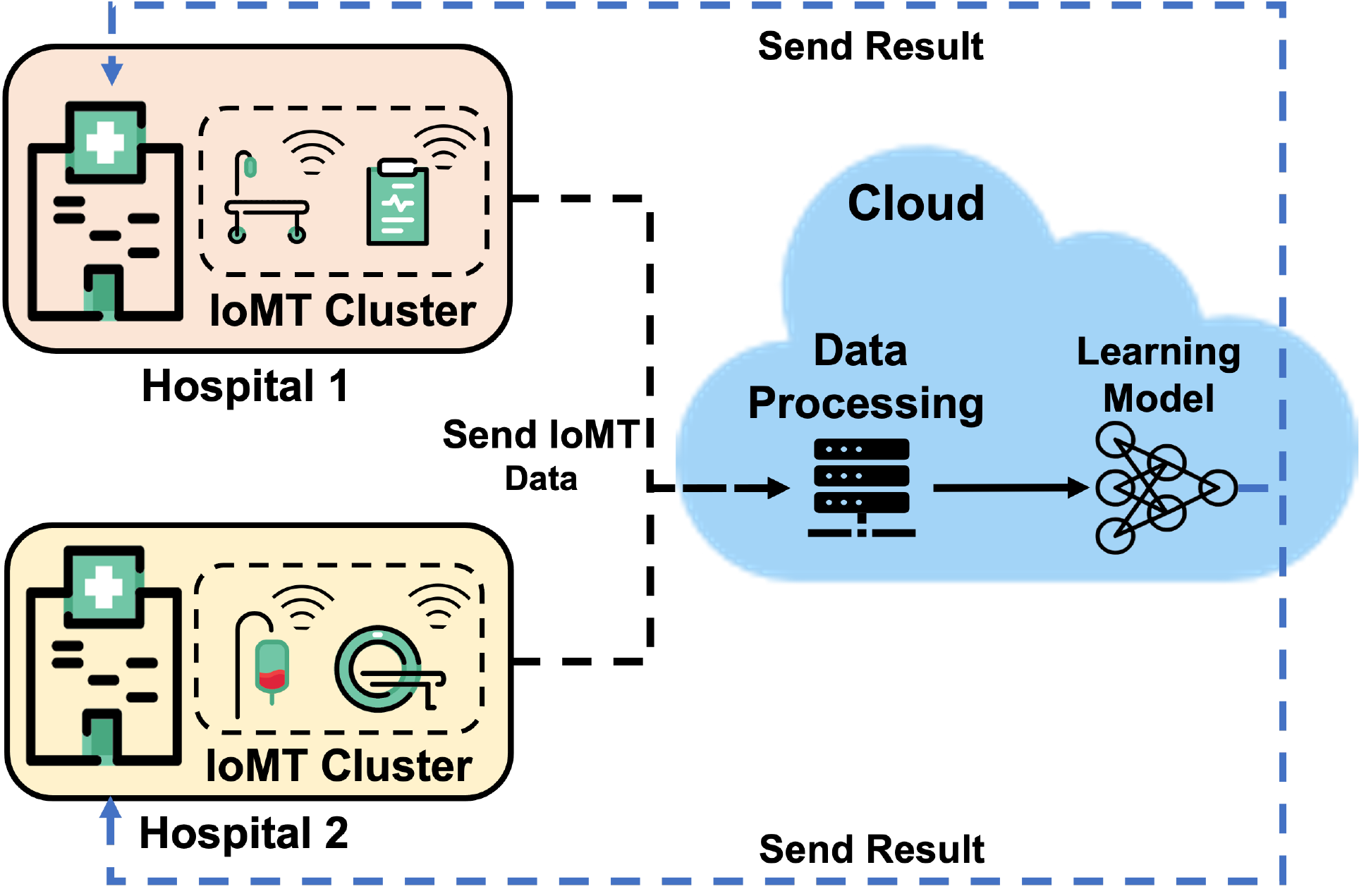}
\caption{\footnotesize{Traditional federated learning application}}
\label{fig:intro}
\end{figure}

As shown in Figure \ref{fig:intro}, cloud computing is one paradigm that emerged to solve the availability of computing and storage resources. Therefore, the cloud is usually used to deploy the DL model for training and data inference. However, sending the raw data from the IoMT cluster to the cloud will be very expensive. This is where edge computing, like edge servers, will be advantageous to process the data before sending it to the cloud. 

It is known that a high-performing Deep Learning (DL) model requires a large and diverse dataset for its training. This large-scale dataset is often obtained from multi-institutional or multi-national data accumulation and voluntary data sharing in the healthcare industry. While massive data collection is essential for the deep learning process, sharing patients' data raises privacy concerns and relative regulations such as the General Data Protection Regulation (GDPR) and Health Insurance Portability and Accountability Act (HIPAA). Due to the rising concerns, healthcare institutions may be prevented from sharing their medical datasets. In some cases where sharing is possible, some restrictions are applied, resulting in inadequate data sharing. 

In recent studies, \cite{li2019convergence} proposed a federated learning model that allows parties to collaboratively train a model by sharing local model updates with a parameter server. Intuitively, this method is safer than centralized training because machine learning models learn from healthcare IoMT data without relying on a third-party cloud to hold their data \cite{yu2021research}. However, federated learning also presents some challenges that may limit its applications in real-world case scenarios. For example, federated learning remains vulnerable to various attacks that may result in leakage of private data \cite{wei2020framework} or poisoned learning model \cite{mothukuri2021survey}. Also, the participants in the current FL setup cannot verify the authenticity of the machine learning model. To protect FL participants' privacy, the existing defense method mainly focuses on ensuring the confidentiality of the machine learning gradients. Differential Privacy (DP) \cite{9253545}, \cite{9325934} is one of the commonly used methods to preserve the privacy of the learning model. Adding DP to a federated learning scenario can improve the privacy of the participants models. However, adding noise into machine learning gradients will reduce the learning model accuracy \cite{9325934}. DP is also ineffective in mitigating poisoning attacks while maintaining model performance resulting in a faulty global model. To tackle the poisoning attack, existing research on anomaly detection \cite{9409113},\cite{9424138} has been explored. However, the existing methods cannot eliminate all the poisoned models and cause the accuracy of the global model to be reduced. Also, they perform the anomaly detection method in a plaintext model. This will lead to another issue where the attacker can perform a parameter stealing attack \cite{8418595} and a membership inference attack \cite{8835245}. Thus, a verifiable and secure anomaly detection method for federated learning scenarios is needed.

This paper proposes a privacy-preserving verification method to eliminate poisoned local models in a federated learning scenario. The proposed method eliminates the compromised  local model while guaranteeing the privacy of the local model's parameters using an SMPC-based encrypted inference process. Once the local model is verified, the verified share of the local model is sent to the blockchain for the aggregation process. SMPC-based aggregation is used to perform the secure aggregation between the blockchain and the hospital. After the aggregation process, the global model is stored in tampered-proof storage. Later, each hospital receives the global model from the blockchain and verifies the authenticity of the global model. The contributions of our work are summarized as follows:
\begin{itemize}
    \item Propose a new blockchain-based federated learning architecture for healthcare systems to ensure the security of the global model used for classifying disease.
    \item Design a privacy-preserving method for local model anomaly detection in a Federated learning scenario with SMPC as the underlying technology. Our encrypted model verification method eliminates the poisoned model while protecting the local model privacy from membership inference attacks and parameter stealing.
    \item Propose an SMPC-based secure aggregation in the blockchain as a platform to decentralize the aggregation process.
    \item We  present a verifiable machine learning model for federated learning participants using blockchain in the IoMT scenario.
\end{itemize}

The rest of this paper is organized as follows. Section \ref{sec:issue} defines the problem and design goals. Section \ref{sec:related} discusses the related work. Then, we present the system architecture and introduce the proposed frameworks in Section \ref{sec:framework}. Next, we describe the experimental setup and evaluation results of the proposed work in Section \ref{sec:exp}. Finally, a conclusion is drawn in Section \ref{sec:con}.

\section{Problem Scenario and Design Goals}\label{sec:issue}

\subsection{Problem Scenario}
\begin{figure}[tbh!]
\centering
\includegraphics[width=1\linewidth]{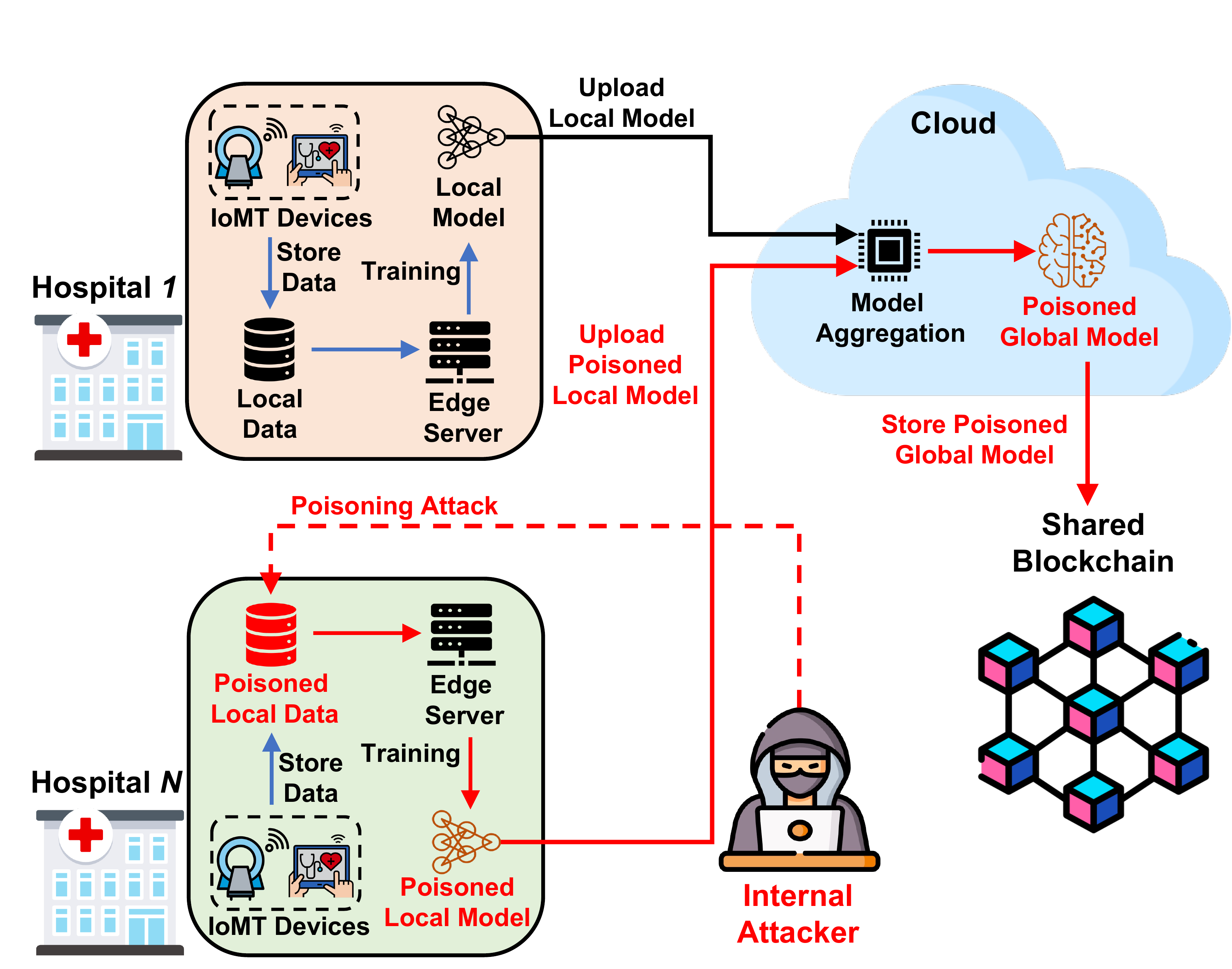}
\caption{\footnotesize{Possible threat in existing federated learning healthcare scenario}}
\label{fig:problem}
\end{figure}

To discuss and highlight the current issues with current federated learning, we use an IoMT-enabled hospital scenario (see Fig. \ref{fig:problem}). Assume that several smart hospitals are placed in different regions with varying patient demographics and diseases. Each smart hospital is equipped with a cluster of IoMT devices. The IoMT devices will be used to scan the patient to detect a severe disease. In The current IoMT scenario, IoMT devices will act as data sources since the IoMT devices are resource-constrained and cannot perform any machine learning algorithm. Hence, each hospital has an edge server with computing resources to execute the machine learning tasks using the local datasets. Nevertheless, due to dataset limitations, the machine learning model accuracy generated from the local datasets is relatively low. Therefore the edge server from each hospital participates in the federated learning platform. In the federated learning platform, locally trained models from the hospital's edge server are collected and aggregated to produce a highly accurate machine learning model without sending private datasets to the cloud provider. Later, the aggregated or global model is sent back to the edge server for another round of federated learning processes. Once the global model reaches the desired accuracy, it will be used to recognize the disease more accurately.

Although the aforementioned federated learning scenario improves the overall machine learning accuracy, it suffers from the following security risks:

\begin{itemize}
    \item Risks of local model security:
    In the current setup of federated learning, every party that sends their local model is sent to the cloud for the aggregation process without checking the model's validity. This traditional FL method introduces the risk of a local model being poisoned. For example, an attacker can perform a poisoning attack and train the model using poisoned data, leading to a faulty local model. Since healthcare data are critical, sending plaintext local models to the cloud can pose privacy risks. Therefore, validating and securing the local model is required to prevent it from various security aspects.
    
    \item Risks of generating a biased aggregated model:
    The model aggregation process of the local model is performed on the cloud services that can be tampered with and produce a biased global model. For example, an attacker can include a poisoned local model during the aggregation process that may lead the global model to have a false classification. Hence, a secure aggregation method is required to encounter the current security problem.
    
    \item Risk of receiving faulty global model:
    In the existing federated learning method, the global model generated from the cloud will be sent back to each edge server in the hospitals. However, the hospital can not verify the global model they received. The attacker can intercept and alter the global model. As a result, the hospital received a faulty global model. From this problem, a global model verification method is required to ensure the integrity of the global model.
    
\end{itemize}

\subsection{Design Goals}
With the risks and threats mentioned above, our goals for preserving privacy in Federated Learning can be decomposed into three aspects as follows:

\begin{itemize}
    
    \item \textbf{Robustness:} 
The proposed work should have the ability to prevent the adversary from poisoning federated learning. This allows the federated learning participant to learn from a benign global model to improve their model accuracy. Also, a robust aggregation method needs to be developed to secure the aggregation process from an attacker.
    
    \item \textbf{Privacy:} 
The prior work \cite{tolpegin2020data} has shown that an attacker can perform a poisoning attack to decrease the global model accuracy by miss-classifying the machine learning model. To protect the federated learning participants, checking the participant's local learning model while maintaining the local model privacy itself is essential.
    
    \item \textbf{Verifiability:} 
The designed method should have the ability to verify the machine learning model, specifically the global model. Since the adversary may alter or poison the global model. In the current federated learning scenario, the participant received the global model from the cloud without knowing the model's authenticity.
    
\end{itemize}

\section{Related Work}\label{sec:related}

Below, we discuss several existing privacy-preservation methods for FL. Then, existing blockchain-based FL will also be presented.

\subsection{Privacy-preserving Federated Learning and Existing Attack}

In FL, data privacy is achieved by sending the model to the client and performing local training. Later, the locally trained model will be collected by the central server and aggregated into a global model. With this method, the participants only shared the local model and did not send any datasets. However, FL itself is not sufficient to provide a privacy guarantee.

Some research has been performed to secure the FL architecture. The author in \cite{9253545} and \cite{9325934} enhance the data privacy in FL with differential privacy (DP) by adding noise in the local datasets. In \cite{9325934}, also anonymize the end-user by adding a proxy server. However, the experiment result show there is a significant accuracy reduction. This privacy-preserving method is unsuitable for FL in healthcare systems since accuracy is essential for the inference process.

Zhang et al. \cite{zhang2020batchcrypt} use fully homomorphic encryption (FHE) to perform aggregation and training processes by performing a batch encryption method. However, all the homomorphic encryption methods are unusable for healthcare scenarios since the training process takes significant time.

Authors in \cite{9194010}, \cite{fang2020local}, and \cite{bagdasaryan2020backdoor} have successfully performed an adversarial attack on FL architecture. The authors have demonstrated a poisoning attack on the local client's datasets. The poisoned model will be generated and impact the global model. Based on the existing attack, DP and FHE method is insufficient against the poisoning attack. 

In \cite{9524709}, the author proposed a privacy-enhanced FL against poisoning adversaries. To secure the machine learning model, they encrypt the model using linear homomorphic encryption. Since they encrypt the model from the first round of FL, the training process will take longer than regular machine learning. After the participants finish the encrypted training process, The local model will send to the server for encrypted aggregation. Based on the results of their experiments, their aggregation method reduces the accuracy of the machine learning model.

Our proposed method performs anomaly detection using an encrypted inference process to eliminate the poisoned local model. Later, we leverage the SMPC-based secure aggregation method. Our secure aggregation method will not affect the accuracy of machine learning. Also, we leverage blockchain for the aggregation process as part of the consensus mechanism to mitigate a single point of failure.

\subsection{Blockchain-based Federated Learning}

Blockchain is known for its immutability and is used for tampered-proof storage. The use of blockchain can track the local or global model for audibility purposes. Combining blockchain with FL can ensure the machine learning model's integrity.

Author in \cite{9285303} proposed verifiable aggregation for FL. Their method follows the concept of blockchain, where they use the hash to compute the digest for verification. Nonetheless, the aggregation and hashing process is performed on a single server. The correct utilization of blockchain technology can overcome the problem.

In tackling the issue, \cite{9019859} proposed decentralized privacy using blockchain-enabled FL. They use blockchain to store and verify the model using cross-validation, but the participant is connected to the same blockchain. In their framework, the participant can use other's local models, which leads to privacy issues.

The work on \cite{9321132} uses a smart contract to verify the global model. The use of smart contracts can audit the authenticity of the global model. However, they did not perform any checks on the local or global model. Also, the local model is not sent to the blockchain, and not possible to perform any audit process. From the proposed work, they can not handle any poisoning attack.

The author in \cite{9420107} closely relates to our work. They proposed a blockchain-based FL for COVID-19 detection using CT imaging. In their proposed work, the local model is sent from the blockchain. For efficiency, they aggregate the model on a single server, leading to a single point of failure and tampering attack. There are no verification processes on the local model before the aggregation. Since it's a shared blockchain, every hospital can access other private local models.

In our proposed framework, we will first check on the local model sent from the hospital through an encrypted inference process. Once it's verified, the local model will be sent to the blockchain. Every blockchain node will receive the verified local models and perform the aggregation process. The global model is sent from the blockchain to the hospital when the consensus is done. With this method, the hospital can verify the authenticity and the model's integrity.

\section{Proposed Framework}\label{sec:framework}
This section presents our proposed blockchain-based federated learning with secure model verification. First, we present an overview of the system architecture. Next, we discuss in detail the various components of our proposed framework. The summary of notations used in the methodology can be seen on Table I.

\begin{table}
\begin{center}
\caption{Notations}
\scalebox{0.7}{
\begin{tabularx}{\linewidth}{@{}XX@{}}
\toprule
    $M_{L}$ & Local Model \\
    $M_{G}$ & Global Model \\
    $M_{Li}^{r+1}$ & Updated Local Model\\
    $V(M_{Ln})$ & Verified Local Model\\
    $HL_{n}$ & Local Image Dataset\\
    $C_{n}$ & IoT Cluster\\
    $S_{n}$ & Edge Server\\
    $B_{n}$ & Blockchain Node\\
\bottomrule
\end{tabularx}
}
\end{center}
\end{table}

\subsection{System Architecture}
\begin{figure}[tbh!]
\centering
\includegraphics[width=1\linewidth]{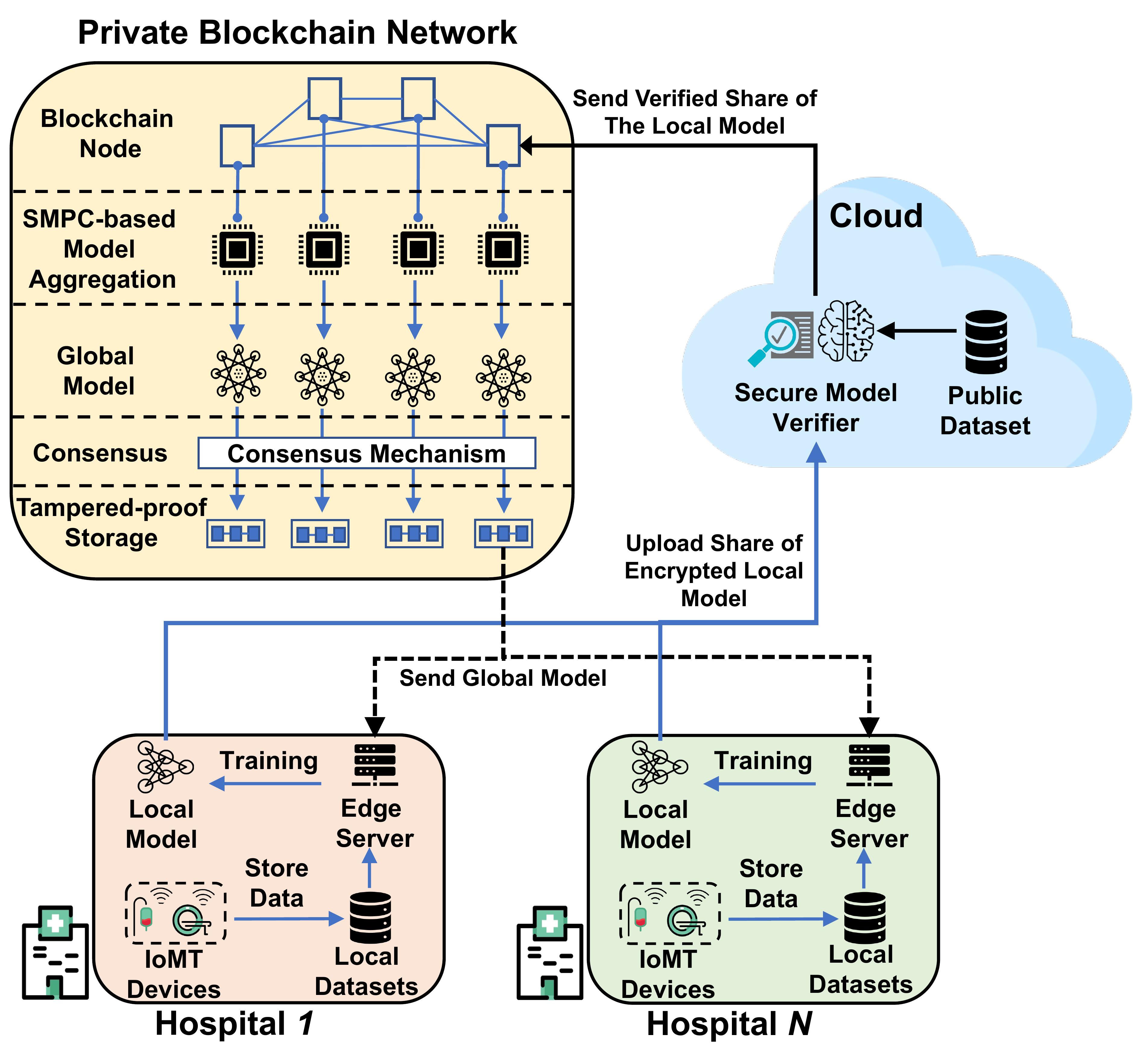}
\caption{\footnotesize{Overview of the proposed framework}}
\label{fig:architecture}
\end{figure}

We propose a verifiable Federated Learning (FL) scenario that leverages SMPC to perform an encrypted local model verification process and secure aggregation on the blockchain node. We assume an $n$ number of hospital $H$, and each hospital has an edge server $S_n$. To maintain the privacy and security of healthcare data, we assume each hospital has its own on-premise server that has enough computing resources to process or train the healthcare data that they have. Every hospital is also equipped with several IoMT devices $Z_n$ for data sources. Because IoMT sensors do not have sufficient computing power to perform a machine learning (ML) algorithm, $S_n$ will be used to execute and run the ML model. As a result, IoMT devices and edge servers in the hospital will form a cluster $C_n(1 \leq n \leq H )$. In the original machine learning, an edge server trains the ML model based on the local dataset and generates a \textit{Local Model} $M_L$. Due to limited datasets from the hospital, the accuracy of the $M_L$ might not be high. Exchanging datasets from another hospital to improve the machine learning accuracy might lead to a privacy issue. To tackle this problem, edge servers from every $C_n$ from each hospital join the cross-silo federated learning environment that involves multiple hospitals with diverse datasets that can produce various $M_L$. Where each hospital (silo) contributes its own datasets to the training process. This can be useful for the hospital that wants to share data for training a model but unable or unwilling to share the raw data itself. In the cross-silo federated learning scenario, every hospital that participates in the FL needs to download the initial \textit{Global Model} denoted as $M_G$ and use it as the based model for the $M_L$ training. Later each hospital sends the trained $M_L$ to an aggregator to aggregate the model. However, a typical FL approaches perform an aggregation process without checking that the $M_L$ is free from any adversarial attack. In the healthcare scenario, securing the $M_L$ from an attacker is essential since the $M_G$ depends on the collected $M_L$. We proposed an encrypted inference process using SMPC to check every $M_L$ before the aggregation. To enhance the privacy of FL, the aggregation process is encrypted and performed in the blockchain node. Later the encrypted global model is stored in tampered-proof storage. Fig. \ref{fig:architecture} gives an overview of the proposed framework.

\subsection{Label Flipping Attack}
The adversarial attack we perform in this paper is a label-flipping attack. With this attack, the adversary's goal is to manipulate the learned parameters of $M$ such that the error is increased for particular source classes. Because this attack is targeted at the error of specific classes, it has increased stealthiness compared to an untargeted attack. The attack is considered successful if the global model incorporates the adversary's malicious updates such that the error for the source classes is increased.

\begin{figure}[tbh!]
\centering
\includegraphics[width=1\linewidth]{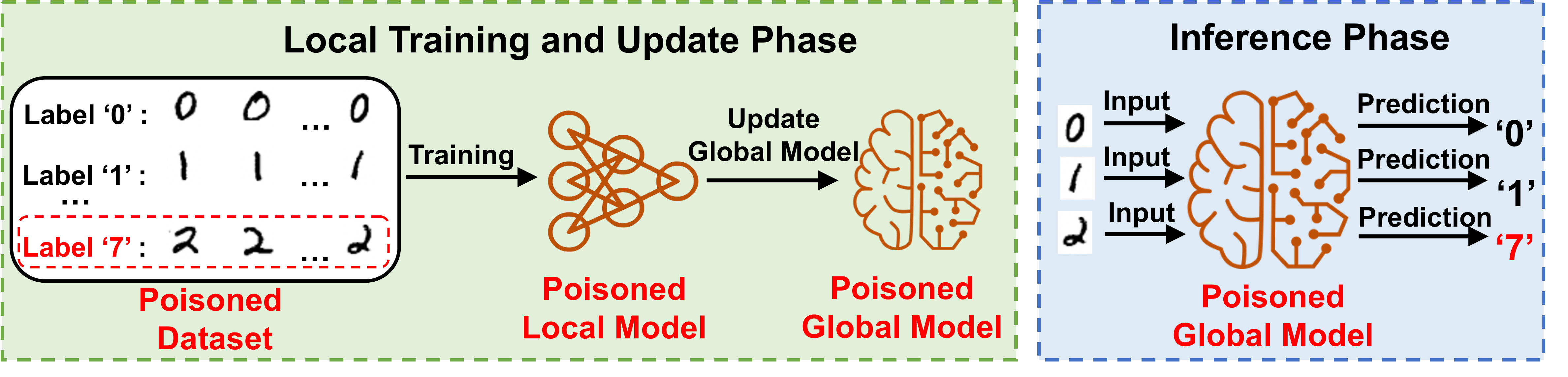}
\caption{\footnotesize{Label Flipping Attack}}
\label{fig:labelflip}
\end{figure}

In our threat model, the participants have complete control over training their local model and can alter the training hyperparameters and process. Malicious participants can use this to their advantage to overcome being in the minority of participants. For example, a malicious participant $P_i$ can scale up their trained model's parameters by a scaling factor before communicating it to the server for aggregation. This can help with poisoning but has to be finely tuned to avoid causing the server to fail to train entirely or prevent the malicious participants poisoning efforts from being detected. This scaling factor can also be adjusted over time to optimize the poisoning rate and evaluation accuracy. The overview of the label-flipping attack is shown in Fig. \ref{fig:labelflip}.

\subsection{Local Model Generation}
\begin{figure}[tbh!]
\centering
\includegraphics[width=1\linewidth]{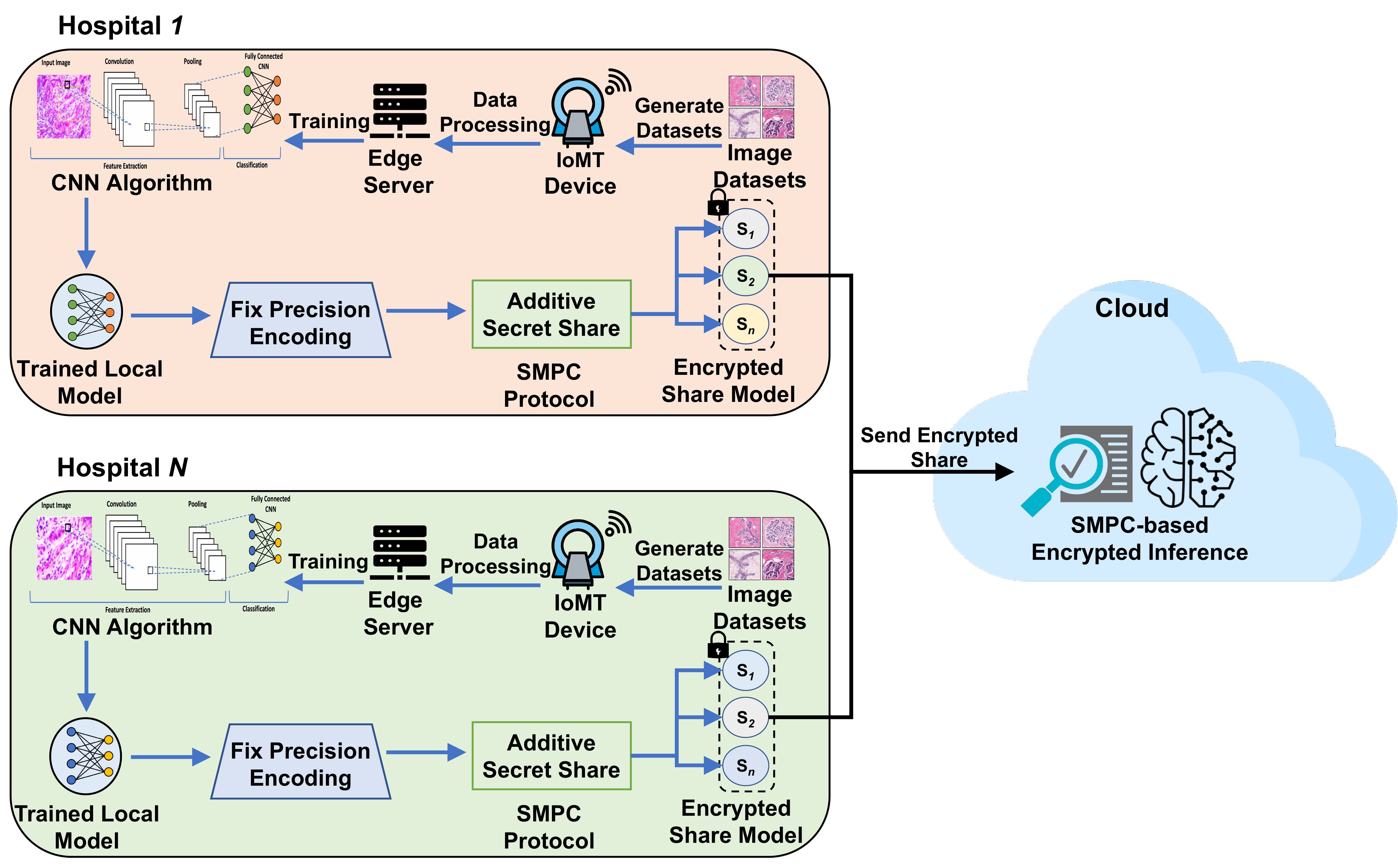}
\caption{\footnotesize{Local Model Generation}}
\label{fig:lmg}
\end{figure}

In this local model generation step, every hospital performs a local model training process using the dataset collected from $C_n$. Fig. \ref{fig:lmg} gives an overview of the local model generation process. In the overview of the proposed framework, the edge server in each hospital receives the training model from the tampered-proof storage. The local model that is being used is a Convolutional Neural Network (CNN)-based image classification. In our scenario, ResNet-18 \cite{he2016deep} is used for deep learning in medical image classification. 

In general, CNN-based image classification takes an input image and classifies it into certain categories of $y$ objects. An edge server $S_n$ has a local dataset $HL_n$ produced from cluster $C_n$. The edge server $S_n$ process the input image as an array of pixels based on the image resolution. For example, the medical image dataset has a set of attribute that needs to be considered during the training process. Based on the medical image dataset, the CNN will see the image's height $h$, width $w$, and dimension $d$. Later, the CNN will read the array of input as follows $h$ x $w$ x $d$. The dimension $d$ is perceived as a three-color channel or RGB in the medical image datasets. A machine learning model that uses CNN works with different layers to train and test the local model. The specific layers that are used in ResNet-18 \cite{he2016deep} consist of \textit{convolutional layers, pooling layers, and fully connected layers}.
At last, the CNN applies \textit{softmax layers} to classify the object with probabilistic between 0 and 1. After $S_n$ performs the local training and testing process with the CNN algorithm, the local model $M_{Ln}$ is generated. The local model will be evaluated for every round in the federated learning setup to achieve a certain accuracy for the global model. In this scenario, an edge server $S_n$ updates the $M_{Ln}$ model using the local datasets $HL_n$ in every federated learning round $r$ as follows:

\begin{equation}
  M_{Ln}^{r+1} = M_G^r - \eta \nabla F(M_G^r,HL_n) 
\end{equation}

Where $M_{Ln}^{r+1}$ denotes the updated local model of client $i$, $M_G^r$ is the current global model, $\eta$ is the local learning rate, $\nabla$ is used to refer to the derivative with respect to every parameter, and $F$ is the loss function. Later, we verify the trained local model through an encrypted inference process to prevent it from a membership inference attack. We leverage SMPC protocol to perform the encrypted inference process. In SMPC protocol, a Trusted Third Party (TTP) provides the necessary variables to keep all the computation in the inference process private. Afterward, the SMPC protocol will encrypt and split the local model into several shares. The encryption and local model splitting will be discussed in the section. \ref{sec:smv}.

\subsection{Secure Model Verification} \label{sec:smv}

\begin{figure}[tbh!]
\centering
\includegraphics[width=1\linewidth]{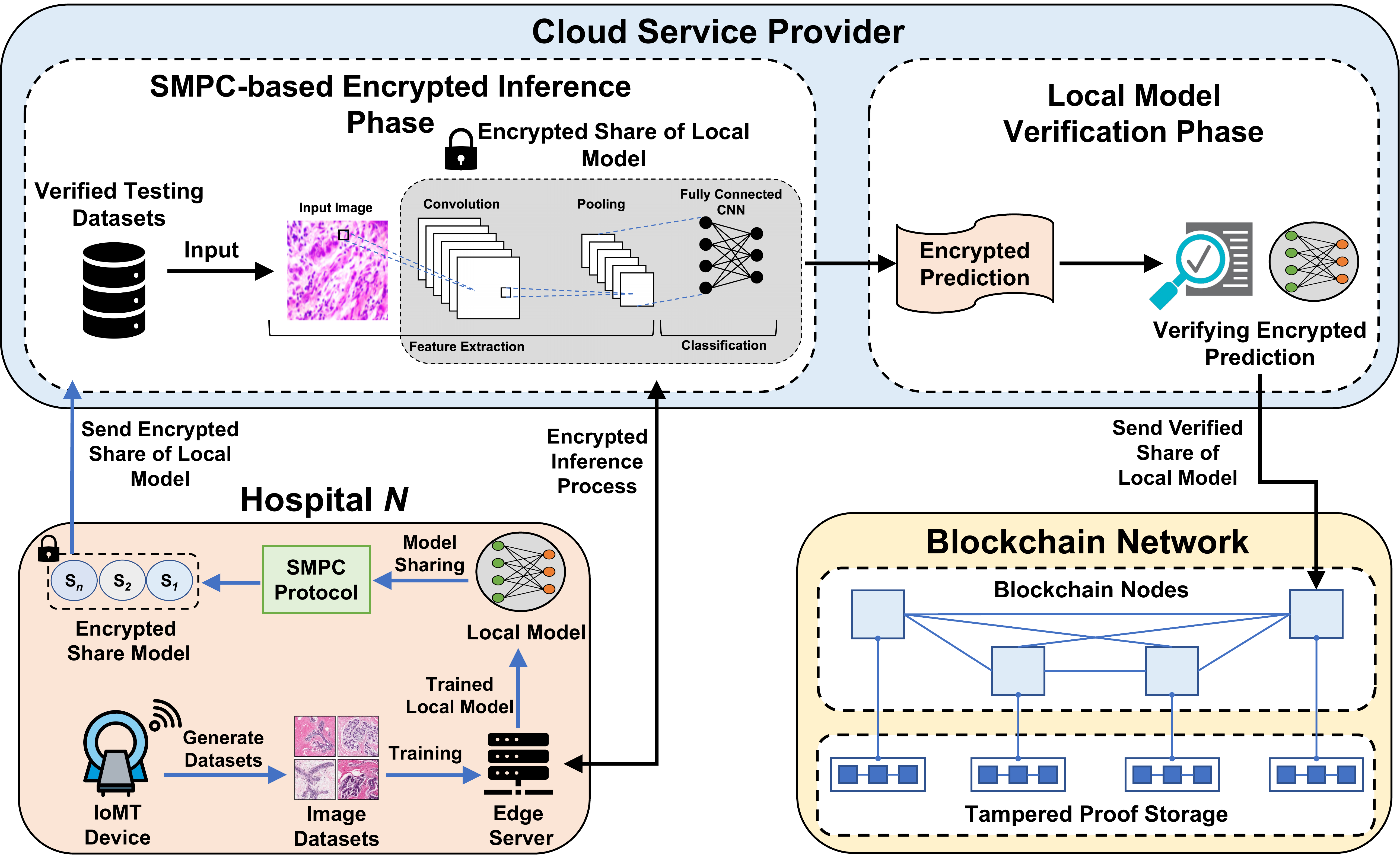}
\caption{\footnotesize{Secure Model Verification}}
\label{fig:smv}
\end{figure}

The secure Model Verification (SMV) phase is performed in the Cloud Service Provider (CSP). SMV leverages secure multi-party computation (SMPC) to perform encrypted inference. In SMPC, we consider Function Secret Sharing (FSS) protocol to allow hospital and CSP to keep their input and model confidential. To achieve this, let $\mathcal{F}$ be a function $f:{0,1}^n\rightarrow G$, where $G$ is an Abelian group. A function share $([[f]]_0, [[f]]_1)$ is generated by $\mathcal{T}$ from $\mathcal{F}$ such that $\mathcal{F}(x) = [[f]]_0(x) + [[f]]_1(x)\ mod\ 2^n$, where $n$ is the number of bits which all values within the computation are encoded into, and $x$ is public input. These function shares are then sent to the hospital $H_n$ and CSP. Suppose a CSP has a private input $y$ from the verified testing datasets to be inferred. To generate a public input $x$, input shares $[y]_0$ and $[y]_1$ are first generated and shared with the hospital $H_n$. Each party first mask these values using a random mask $[\alpha]$. This is done by computing $[y]_0 + [\alpha]_0$ and $[y]_1 + [\alpha]_1$. Finally, we obtain $x$ by computing $x = y + \alpha$. By applying $x$ to function shares $[[f]]_0$ and $[[f]]_1$, we get output shares from each party, which can be used to reconstruct the output. Later, the testing output will be compared with the CSP threshold to determine whether the local is being compromised or not. The overview of the SMV can be seen in Fig. \ref{fig:smv}.

In our scenario, during the encrypted inference, the SMPC needs to encrypt and create several shares from the local model $M_{Ln}$. Suppose a set of input images $Y=\{y_1,y_2,...,y_n\}$ represents a testing image from the verified testing datasets in CSP. The verified $Y$ will be used to test the local model $M_{Ln}$ by performing an encrypted inference process. All basic operations to infer an input $Y$ on a $M_{Ln}$ follow additive secret sharing workflow. The steps can be seen in Algorithm \ref{alg: smv}. First, input $y_n \in Y$ needs to be converted to integer values. This is done by using \textit{fix\_precision()} function. Then, $Y$ is encrypted by generating two shares, $[y]_0$ and $[y]_1$. Each share is sent to the hospital $H_n$ and CSP. The address of the share is then saved in $Enc\_data$. Each party then masks their share using random mask $\alpha_{i}, i \in {0,1}$, where $i$ is the CSP and hospital id. This $\alpha$ is provided by the $\mathcal{T}$. The masked value is then added together to generate public value $x$. Next, $M_{Ln}$ is encrypted using the FSS protocol. This FSS protocol produces function shares $[[f]]_0$ and $[[f]]_1$ for each machine learning operation and is distributed to the CSP and hospital. The function shares address is stored in $Enc\_model$. $y_n \in Y$ is then fed to the $Enc\_model$ to produce shared output. Finally, the shared output is reconstructed to produce a set of final output $\mathcal{O} = \mathcal{O}_1, \mathcal{O}_2,..., \mathcal{O}_n$. Later, a set of $\mathcal{O}$ will be compared with the threshold that CSP has to determine whether the local model is faulty or not. Once the local model $M_{Ln}$ passes the threshold requirement, the CSP will send the shared of the verified share of the local model $VS_n(M_Ln)$ to the blockchain network for the aggregation process.

\begin{algorithm}[ht!]
\SetAlgoNoLine
\caption{\footnotesize{Secure Model Verification}}
\label{alg: smv}
\KwIn
{
    \begin{minipage}[t]{10cm}%
     \strut
        $M_{Ln}$ - Local Model
        
        $Y=\{y_1,y_2,...,y_n\}$ - Set of Verified Testing Images
     \strut
    \end{minipage}
}
\KwOut{
    \begin{minipage}[t]{10cm}%
     \strut
        $\mathcal{O} = \{\mathcal{O}_1, \mathcal{O}_2,..., \mathcal{O}_n\}$ - Set of Inference Output
     \strut
    \end{minipage}
}
\ForEach{$y_n \in Y$}
    {$Y = Y.fix\_precision()$
    
    $Enc\_data = Y.encrypt()$
    
    $Enc\_model = M_Ln.encrypt()$
    
    $Enc_prediction = Enc\_model(Enc\_data)$
    
    $\mathcal{O} = Enc_prediction.decrypt()$
   }
    
    \textbf{return} $\mathcal{O} = \{\mathcal{O}_1, \mathcal{O}_2,..., \mathcal{O}_n\}$
    
    \textbf{end}

\end{algorithm}

\subsection{Blockchain-based Secure Aggregation}

After the CSP performs a secure model verification on the local model $M_{Ln}$ and is verified, CSP sends the $VS_n(M_Ln)$ to the blockchain node $B_n$ for secure aggregation. At this stage, only the blockchain node $B_n$ and the hospital hold the encrypted share of the local model. Later, $B_n$ and $S_n$ from the respective hospitals will perform the secure aggregation process to generate the global model $M_G$. For the secure aggregation process, we leverage an additive secret-sharing scheme. Additive secret sharing \cite{doganay2008distributed} allows a trusted party $\mathcal{T}$ to share a secret $s$ among $n$ parties $P_1, P_2, ... P_n$, such that to reveal $s$, $n$ node must share their secret. This process starts with a high number of prime number $Q$ generation. Then, $s$ is split into $n$ number of shares ${s_1, s_2, ..., s_n}$. In this scheme, the shares of $s$ must satisfy that 

$$s = \left(\sum_{i=1}^{n} s_i\right)\ mod\ Q$$. 

This can be done by choosing $s_1, s_2, ..., s_{n-1} \in [0, Q-1]$, and $s_n = \left(s - \left(\sum_{i=1}^{n-1} s_i\right)\right)\ mod\ Q$. The reconstruction of $s$ can be done by calculating $s = \left(\sum_{i=1}^{n} s_i\right)\ mod\ Q$. For the aggregation process, each blockchain node $B_n$ are able to receive multiple set of verified share model $VS_n(M_Ln)=\{VS_1(M_{L1}),VS_2(M_{L2}),\dots,VS_i(M_{Li})\}$ from different hospital. By having a share of the local model, $B_n$ can not reconstruct or achieve any private information from the model.

In additive secret sharing, properties such as addition, subtraction, and multiplication are supported. Suppose that $B_1$ has a secret $s$ and $S_1$ has a secret $u$. Additionally, there exists a $\mathcal{T}$. In a particular case, $B_1$ and $S_2$ want to know the sum of their secrets without revealing the true value of their secrets. To calculate the sum, the protocol simply shares the secrets of the two parties into $n$ number of shares. Particularly, $s$ is divided into three shares $(s_1, s_2, s_3)$ and $u$ into another shares $(u_1, u_2, u_3)$. Each party will hold one share of each secret. For example, $B_1$ holds ($s_1$, $u_1$), $S_2$ holds ($s_2$, $u_2$), and $\mathcal{T}$ holds ($s_3$, $u_3$). Then, to calculate the sum of the two secrets, each party adds up the shares they hold using $m_n = (u_n + s_n)\ mod\ Q$, where $m_n$ is the sum of shares that party $n$ holds. Finally, the sum of shares is computed as follows: 

$$ u + s =  \left(\sum_{i=1}^{n} m_i\right)\ mod\ Q$$

For Secure federated averaging aggregation proposed in \cite{ziller2021pysyft}, an addition operation is employed to obtain the model parameters average. In our scenario, we assume there are $n$ number of blockchain nodes $B_n$ and a trusted party $\mathcal{T}$. Suppose that we have a verified local model $VS_n(M_{Ln})$ from $B_{n}$. The aggregation process is defined in Algorithm \ref{alg: secAgg}. 

To begin, a trusted third party $\mathcal{T}$ computes $Q$, which is visible to $B_n$ and $S_n$. Then, it initializes a model $M_G$ to store the aggregated model parameters. Each $S_n$ creates $n$ shares of their parameters ($param$). Suppose the model parameters have a length of $J$ and $j \in [0, J]$. For each parameter of participants models $param[j]$, $S_n$ create $n$ shares of the local model and send it to blockchain node $B_n$. However, since additive secret sharing takes a finite abelian group, any parameter with a floating number must be converted to an integer. Hence, using the $fixed\_precision()$ function, floating points are truncated to the fourth decimal. A particular share of parameter $param[j]$ from each party that joins the computation $P_i$ is denoted as $share\_i\_j$. To calculate a model parameter average, each $P_i$ computes the sum of shares on $param[j]$ as follows: 

$$P_{i\_s} = \left(\sum_{i = 1}^{n} share\_i\_j\right)\ mod\ Q$$

where $i$ is the participant id, $n$ is the total number of participants, and $j$ is the parameter index. The sum of shares from all $P_i$ are then added up together and averaged as follows:

$$M_{G}.param[j] = \frac{\left(\sum_{i = 1}^{n} P_{i\_s}\right)\ mod\ Q}{n}$$

The above equation results in an aggregated result of a parameter from all hospitals' trained local models. The result is then used to replace $param[j]$ in ${M}_{G}$. Finally, an aggregated model $M_G$ is produced after the processes are iterated through all of the model parameters.

\begin{algorithm}[ht!]
\SetAlgoNoLine
\caption{\footnotesize{Blockchain-based Secure Aggregation }}
\label{alg: secAgg}
\KwIn
{
    \begin{minipage}[t]{10cm}%
     \strut
        $n$ - Number of parties
        
        $\mathcal{M}\_p\_len$ - Model's parameters length
     \strut
    \end{minipage}
}
\KwOut{
    \begin{minipage}[t]{10cm}%
     \strut
        $M_G$ - Aggregated model
     \strut
    \end{minipage}
}
    % \Init{}{
    \textbf{Initialization:}
    
    $\mathcal{T}$ initialize a high prime number, $Q$
    
    initialize aggregated model, $M_G$
    
    \textbf{begin}
    
    \For{$j \in range(0, \mathcal{M}\_p\_len)$}
        {
            
            \Foreach{$P_i \in P_{i-1}$}
                {   
                    
                    $\mathcal{M}_{i} = P_i.model()$
                    
                    $p = \mathcal{M}_{i}.param[j]$
                    
                    $p.fixed\_precision()$
                    
                    $shares = 0$
                    
                    \While{$n \neq 0$}{ 
                    
                        \eIf{$n = 1$}{
                
                            $share\_i\_j = (p - shares)\ mod \ Q$
                            
                            $send(share\_i\_j, P_i, idx)$
                            
                        }{
                        
                            $share\_i\_j = random(0, Q-1)$
                            
                            $send(share\_i\_j, P_i, idx)$
                            
                            $shares = shares + share\_i\_j$
                            
                        }
                        
                        $n--$
                    }
                    
                }
                
            $M_G.param[j] = 0$
            
            \Foreach{$P_n \in P_{i}$}{
            
                $P_{i\_s} = \left(\sum_{i = 1}^{n} share\_i\_j\right)\ mod\ Q$
            
            }
            
            $M_G.param[j] = \frac{\sum_{i = 1}^{n} P_{i\_s}\ mod\ Q}{n}$
            
        }
      
    \textbf{return} $M_G$
    
    \textbf{end}

\end{algorithm}

After the global model $M_G$ is generated, the blockchain node $B_n$ runs a consensus mechanism. The consensus mechanism verifies the global model produced by the $B_n$. If the majority of hashes of corresponding models are the same, the $B_n$ in the blockchain network adds the global model $M_G$ as a block in the blockchain or the tampered-proof storage. Later the global model is sent to all edge servers $S_n$ as the update of the federated learning rounds.

\section{Results and Discussion}\label{sec:exp}

In this section, we show several experiments conducted to evaluate the performance of our proposed framework. Experimental setup, dataset, and CNN model are discussed in Section \ref{sec:setup} and \ref{sec:data}, respectively. Section \ref{sec:results}, discuss the experimental results.   

\subsection{Experimental Set-up}\label{sec:setup}

In our experiments, we run the aggregation node and the hospital server with the AWS EC2 cloud. Since the training process requires considerable computing power, we use \textit{P3} machine. Instance \textit{ml.p3.8xlarge}, has 4 NVIDIA Tesla V100 with 64 GB memory that has Peer to Peer connection between the GPU. This machine has 32 vCPUs and 244 GB of RAM. 

For cost and performance efficiency, AWS EC2 provides \textit{G4DN} series. The \textit{G4DN} instance from AWS runs the application on a virtual CPU and is optimized for machine learning inference and small-scale training. We use \textit{g4dn.12xlarge} series with 48 virtual CPUs, 192 GB Memory, and 4 NVIDIA T4 GPUs.

For the blockchain implementation, we develop our private blockchain with Python programming language \cite{van2007python} and leverage proof of work as the consensus mechanism. We deploy our customized private blockchain in AWS EC2 \textit{t2.2xlarge} instance. Inside the VM, our python code generated several virtual blockchain nodes that connect with a peer-to-peer connection. The customization in the private blockchain is required to support the secure aggregation that leverages in this paper. The federated learning application is developed using PyTorch \cite{paszke2019pytorch}. 

\subsection{Dataset and Model}\label{sec:data}

We use a dataset from Medical MNIST (MedMNIST) \cite{9434062} for the experiments. These datasets are commonly used for benchmarking in the machine learning framework. Therefore, we have used them to evaluate the performance of our proposed approach. The proposed FL-based approach uses the dataset to train and test the local model on the client side. For all our experiments, we split the training and testing sets. We evenly distributed the training and test sets among the participants based on the number of participants.

MedMNIST is a collection of standardized biomedical images consisting of 12 datasets. The MedMNIST dataset is designed to perform classification on lightweight images with various data scales and diverse tasks (e.g., multi-class and multi-label). All images are pre-processed into 28 x 28 with the corresponding classification labels. 

From MedMNIST, we choose two specific datasets: TissueMNIST and OCTMNIST. We choose these two specific datasets because they have more than 100.000 samples.

TissueMNIST contains 236,386 human kidney cortex cell samples, segmented from 3 reference tissue specimens and organized into eight categories. The TissueMNIST samples are split with a ratio of 7 : 1 : 2 into training, validation, and test set. 

OCTMNIST contains 109,309 valid optical coherence tomography (OCT) images for retinal diseases. The OCTMNIST dataset comprises four diagnosis categories, leading to a multi-class classification task. OCTMNIST samples are split with a ratio of 9 : 1 into training and validation sets and use its source validation set as the test set. 

Both source images are gray-scale, and their sizes are 28 x 28. The exact number of the sample distribution can be seen in Table. \ref{tab1}. Since we are working on the healthcare FL scenario, the dataset is in Table. \ref{tab1} will be divided evenly among the client. In our experiments, we have ten different hospitals as a client.

We use ResNet 18 \cite{he2016deep} for the machine learning model. ResNet 18 is a convolutional neural network (CNN) model that has 18 layers deep and about 11M parameters; it can also load a pre-trained version of the trained model. Since ResNet 18 has three input channels, we convert gray-scale images into RGB images. For the training process, we set the batch size to 64. We utilize an Adam optimizer \cite{kingma2014adam} with an initial learning rate of 0.001 and train the model for 60 epochs.

%%%%
\begin{table}[h]
\begin{center}
%\scalebox{0.7}{
\resizebox{\columnwidth}{!}{
    \begin{tabular}{ |c|c|c|c|c| }
        \hline
        \textbf{Dataset} & \textbf{Total Samples} & \textbf{Training} & \textbf{Validation} & \textbf{Testing}\\
        
        \hline
        TissueMNIST \cite{9434062} & 236,386 & 165,466 & 23,640 & 47,280\\
        \hline
        OCTMNIST \cite{9434062} & 109,309 & 97,477 & 10,832 & 1,000\\
        \hline

        \hline
    \end{tabular}
    }
    \end{center}
    \caption{\footnotesize{Samples distribution}}
    \label{tab1}
\end{table}
%%%%

\subsection{Result Analysis}\label{sec:results}

We perform an FL training with ResNet 18 as a local model for the first result analysis. In comparison, we evaluate our FL architecture with OCTMNIST and TissueMNIST. The dataset is already spread evenly amongst ten clients. From the training accuracy, OCTMNIST reaches 92\% and TissueMNIST has 80\% accuracy. The training accuracy can still be increased if we run more epochs because TissueMNIST has more than 200.000 sample data. For a fair comparison and to avoid overfitting problems with another dataset, we run the experiments with two epochs in local clients within 25 rounds. Based on the result, the model accuracy for both datasets converges after 35 epochs.

%%%%exp1
\begin{figure}[tbh!]
\centering
\begin{subfigure}[tbh!]{0.45\columnwidth}
    \resizebox{1\columnwidth}{!}
    {    
    \begin{tikzpicture}
                    \begin{axis}[
                    xlabel={Training Epoch},
                    ylabel={Accuracy},
                    symbolic x coords = {5,10,15,20,25,30,35,40,45,50},
                    xticklabel style={anchor= east,rotate=45 },
                    xtick=data,
                    ymax=80,
                    ymin=0,
                    legend pos=south east,
                    ymajorgrids=true,
                    grid style=dashed,
                    legend style={nodes={scale=1, transform shape}},
                    label style={font=\Large},
                    tick label style={font=\Large}
                ]
                \addplot+[mark size=3pt]
                    coordinates {
                    (5, 58.316544)
                    (10, 60.453211)
                    (15, 65.17625)
                    (20, 67.9879)
                    (25, 70.87)
                    (30, 71.657886)
                    (35, 72.1)
                    (40, 72.2)
                    (45,72.3)
                    (50,72.4)

                    };
                \addplot+[mark size=3pt]
                    coordinates {
                    (5, 35.5348)
                    (10, 37.55492)
                    (15, 40.98864)
                    (20, 44.34276)
                    (25, 45.0)
                    (30, 49.65745)
                    (35,50)
                    (40,52)
                    (45,53)
                    (50,54)
                    
                    };
                \addplot+[mark size=3pt]
                    coordinates {
                    (5, 36.2871)
                    (10, 36.9437)
                    (15, 38.2653)
                    (20, 40.1876)
                    (25, 43.2765)
                    (30, 47.7653)
                    (35, 49)
                    (40, 49)
                    (45, 50)
                    (50, 51)            
                    };
                    \legend{FL Global Model, Central ML Model, Average FL Clients}
                    \end{axis}
            \end{tikzpicture}
    }
    \caption{\tiny{OCTMNIST}}
    \label{exp1a}
\end{subfigure}
~
~
~
\begin{subfigure}[tbh!]{0.45\columnwidth}
    \resizebox{1\columnwidth}{!}
    {
        \begin{tikzpicture}
                    \begin{axis}[
                    xlabel={Training Epoch},
                    ylabel={Accuracy},
                    symbolic x coords = {5,10,15,20,25,30,35,40,45,50},
                    xticklabel style={anchor= east,rotate=45 },
                    xtick=data,
                    ymax=80,
                    ymin=0,
                    legend pos= south east,
                    ymajorgrids=true,
                    grid style=dashed,
                    legend style={nodes={scale=1, transform shape}},
                    label style={font=\Large},
                    tick label style={font=\Large}
                ]
                \addplot+[mark size=3pt]
                coordinates {
                    (5, 58.2349934)
                    (10, 59.873409746)
                    (15, 61.953733)
                    (20, 63.64883034)
                    (25, 65.66728192)
                    (30, 65.908723)
                    (35, 66)
                    (40, 67.2)
                    (45, 67.3)
                    (50, 67.5)
                    };
                \addplot+[mark size=3pt]
                    coordinates {
                    (5, 45.57589823)
                    (10, 48.2656323)
                    (15, 50.232876542)
                    (20, 52.2763923)
                    (25, 53.08776253)
                    (30, 55.6253728)
                    (35, 55.8)
                    (40, 56.6)
                    (45, 57.2)
                    (50, 57.8)
                    };
                \addplot+[mark size=3pt]
                    coordinates {
                    %(1, 44.357923)
                    (5, 46.122344)
                    (10, 47.563479937)
                    (15, 50.343782364)
                    (20, 51.34783949)
                    (25, 52.34798983)
                    (30, 54.78366778)
                    (35, 55)
                    (40, 55.4)
                    (45, 55.8)
                    (50, 56)
                    };
                   \legend{FL Global Model, Central ML Model, Average FL Clients}
                    \end{axis}
            \end{tikzpicture}
    }
    \caption{\tiny{TissueMNIST}}
    \label{exp1b}
\end{subfigure}
\caption{\footnotesize{Evaluation comparison between FL architecture and single deep learning model using OCTMNIST and TissueMNIST}}
\label{exp1}
\end{figure}
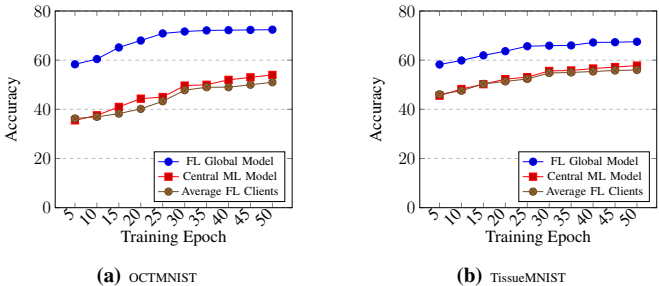
%%%%exp1

In Fig. \ref{exp1}, we compare our FL architecture with the single deep learning model. In this experiment, we replicate the experiment from \cite{9434062} and perform the training process using 60 epochs. In Fig. \ref{exp1a} and Fig. \ref{exp1b}, we show the evaluation accuracy of the global model, single ML model, and the average of the FL participants. In OCTMNIST (Fig.\ref{exp1a}), the average client's accuracy and the centralized training can reach 50\% accuracy. However, the global model evaluation has 72\% accuracy in the same epoch. In this case, the FL architecture can increase 22\% evaluation accuracy under the same setup. In TissueMNIST (Fig. \ref{exp1b}), the global model evaluation accuracy can increase the overall model up to 10\% compared to the single ML model. From Fig. \ref{exp1}, we can see that FL architecture can significantly improve the model evaluation accuracy.

%%%%exp2
\begin{figure}[tbh!]
\centering
\begin{subfigure}[tbh!]{0.45\columnwidth}
    \resizebox{1\columnwidth}{!}{
    \begin{tikzpicture}
                    \begin{axis}[
                    xlabel={Training Epoch},
                    ylabel={Accuracy},
                    symbolic x coords = {10,15,20,25,30,35,40,45,50},
                    xticklabel style={anchor= east,rotate=45 },
                    xtick=data,
                    ymax=80,
                    ymin=0,
                    legend pos=south east,
                    ymajorgrids=true,
                    grid style=dashed,
                    legend style={nodes={scale=1, transform shape}},
                    label style={font=\Large},
                    tick label style={font=\Large}
                ]
                \addplot+[mark size=3pt]
                    coordinates {
(10, 45.3125)
(15, 50.0)
(20, 53.90625)
(25, 50.78125)
(30, 56.25)
(35, 57)
(40, 57)
(45, 58)
(50, 58)

                    };
                \addplot+[mark size=3pt]
                    coordinates {
(10, 53.125)
(15, 53.90625)
(20, 46.875)
(25, 56.25)
(30, 57.03125)
(35, 57)
(40, 57)
(45, 58)
(50, 58)
                    };
                \addplot+[mark size=3pt]
                    coordinates {
(10, 46.875)
(15, 50.0)
(20, 50.0)
(25, 51.5625)
(30, 50.78125)
(35, 50)
(40, 51)
(45, 52)
(50, 52)
                    };
                \addplot+[mark size=3pt]
                    coordinates {
(10, 46.09375)
(15, 46.875)
(20, 44.53125)
(25, 50.0)
(30, 50.78125)
(35, 50)
(40, 51)
(45, 51)
(50, 51.5)
                    };
                \addplot+[mark size=3pt]
                    coordinates {
(10, 52.34375)
(15, 52.34375)
(20, 52.34375)
(25, 53.90625)
(30, 53.125)
(35, 53)
(40, 54)
(45, 54)
(50, 54)
                    };
                \addplot+[mark size=3pt]
                    coordinates {
(10, 50.78125)
(15, 50.0)
(20, 50.0)
(25, 51.5625)
(30, 50.78125)
(35, 51)
(40, 51)
(45, 52)
(50, 52)
                    };
                \addplot+[mark size=3pt]
                    coordinates {
(10, 43.75)
(15, 43.75)
(20, 47.65625)
(25, 43.75)
(30, 52.34375)
(35, 52.4)
(40, 52.5)
(45, 52.8)
(50, 58.9)
                    };
                \addplot+[mark size=3pt]
                    coordinates {
(10, 46.875)
(15, 47.65625)
(20, 50.0)
(25, 50.78125)
(30, 51.5625)
(35, 51)
(40, 52)
(45, 52)
(50, 52)
                    };
                \addplot+[mark size=3pt]
                    coordinates {
(10, 46.09375)
(15, 51.5625)
(20, 51.5625)
(25, 57.03125)
(30, 55.46875)
(35, 55.8)
(40, 55.8)
(45, 55.9)
(50, 55.9)
                    };
                \addplot+[mark size=3pt]
                    coordinates {
(10, 33.59375)
(15, 36.71875)
(20, 35.9375)
(25, 36.71875)
(30, 39.0625)
(35, 39.2)
(40, 39.5)
(45, 39.8)
(50, 40)
                    };
                \addplot+[mark size=3pt, color = green]
                    coordinates {
(10, 57.5)
(15, 62)
(20, 62)
(25, 63)
(30, 67)
(35, 67)
(40, 67)
(45, 68)
(50, 68)
                    }; 
                \end{axis}
            \end{tikzpicture}      
    }
    \caption{\tiny{OCTMNIST - 10\% malicious Clients}}
    \label{exp2a}
\end{subfigure}
~
~
~
\begin{subfigure}[tbh!]{0.45\columnwidth}
    \resizebox{1\columnwidth}{!}{
        \begin{tikzpicture}
                    \begin{axis}[
                    xlabel={Training Epoch},
                    ylabel={Accuracy},
                    symbolic x coords = {10,15,20,25,30,35,40,45,50},
                    xticklabel style={anchor= east,rotate=45 },
                    xtick=data,
                    ymax=80,
                    ymin=0,
                    legend pos=north west,
                    ymajorgrids=true,
                    grid style=dashed,
                    legend style={nodes={scale=1, transform shape}},
                    label style={font=\Large},
                    tick label style={font=\Large}
                ]
                \addplot+[mark size=3pt]
                    coordinates {
(10, 56.9679054054054)
(15, 59.35388513513513)
(20, 59.079391891891895)
(25, 60.34628378378378)
(30, 59.90287162162162)
(35, 59)
(40, 60)
(45, 60.3)
(50, 60.9)
                    };
                \addplot+[mark size=3pt]
                    coordinates {
(10, 55.13091216216216)
(15, 58.50929054054054)
(20, 59.016047297297305)
(25, 59.94510135135135)
(30, 59.88175675675676)
(35, 59)
(40, 60)
(45, 60.5)
(50, 60.8)
                    };
                \addplot+[mark size=3pt]
                    coordinates {
(10, 55.173141891891895)
(15, 57.60135135135135)
(20, 57.64358108108109)
(25, 59.501689189189186)
(30, 59.26942567567568)
(35, 59)
(40, 60)
(45, 60)
(50, 59.5)
                    };
                \addplot+[mark size=3pt]
                    coordinates {
(10, 56.376689189189186)
(15, 60.17736486486487)
(20, 59.860641891891895)
(25, 60.8741554054054)
(30, 60.66300675675676)
(35, 59)
(40, 60)
(45, 60)
(50, 60.5)
                    };
                \addplot+[mark size=3pt]
                    coordinates {
(10, 55.65878378378378)
(15, 58.467060810810814)
(20, 58.06587837837838)
(25, 60.0929054054054)
(30, 59.501689189189186)
(35, 60)
(40, 60.4)
(45, 60)
(50, 60.9)
                    };
                \addplot+[mark size=3pt]
                    coordinates {
(10, 57.05236486486487)
(15, 59.45945945945946)
(20, 60.70523648648649)
(25, 61.61317567567568)
(30, 61.570945945945944)
(35, 60.5)
(40, 60.8)
(45, 61)
(50, 61)
                    };
                \addplot+[mark size=3pt]
                    coordinates {
(10, 56.08108108108109)
(15, 57.8758445945946)
(20, 59.64949324324324)
(25, 59.755067567567565)
(30, 60.17736486486487)
(35, 59)
(40, 59)
(45, 60)
(50, 60)
                    };
                \addplot+[mark size=3pt]
                    coordinates {
(10, 55.4054054054054)
(15, 59.90287162162162)
(20, 59.776182432432435)
(25, 60.641891891891895)
(30, 60.43074324324324)
(35, 59)
(40, 60)
(45, 60.3)
(50, 60.5)
                    };
                \addplot+[mark size=3pt]
                    coordinates {
(10, 56.735641891891895)
(15, 58.445945945945944)
(20, 59.079391891891895)
(25, 59.64949324324324)
(30, 59.248310810810814)
(35, 60)
(40, 60)
(45, 60.4)
(50, 60.5)
                    };
                \addplot+[mark size=3pt]
                    coordinates {
(10, 31.798986486486484)
(15, 35.557432432432435)
(20, 36.592060810810814)
(25, 37.5)
(30, 37.880067567567565)
(35, 37.8)
(40, 38)
(45, 38)
(50, 38)
                    };
                \addplot+[mark size=3pt, color = green]
                    coordinates {
(10, 56)
(15, 58)
(20, 59)
(25, 60)
(30, 60)
(35, 61)
(40, 61)
(45, 61)
(50, 61)
                    }; 
                \end{axis}
            \end{tikzpicture}  
    }
    \caption{\tiny{TissueMNIST - 10\% malicious Clients}}
    \label{exp2b}
\end{subfigure}
\caption{\footnotesize{Poisoning attack on FL architecture with 10\% malicious clients on OCTMNIST and TissueMNIST. The plot show the accuracies of benign 1 \ref{5c1}, benign 2 \ref{5c2}, benign 3 \ref{5c3}, benign 4 \ref{5c4}, benign 5 \ref{5c5}, benign 6 \ref{5c6}, benign 7 \ref{5c7}, benign 8 \ref{5c8}, benign 9 \ref{5c9}, malicious 10 \ref{5c10}, and, global evaluation \ref{5gm}.}}
\label{exp2}
\end{figure}
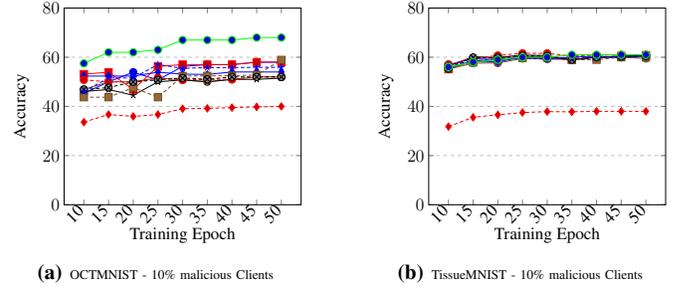
%%%%exp2

%%%%exp3
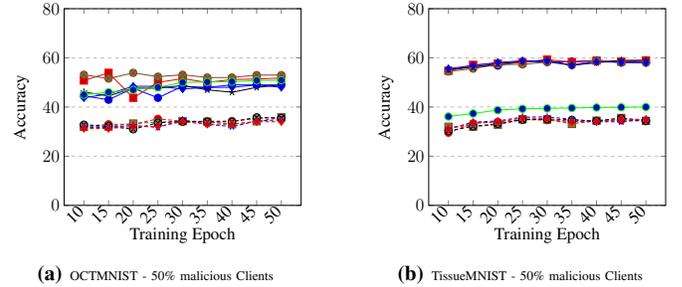
\begin{figure}[tbh!]
\centering
\begin{subfigure}[tbh!]{0.45\columnwidth}
    \resizebox{1\columnwidth}{!}{
    \begin{tikzpicture}
                    \begin{axis}[
                    xlabel={Training Epoch},
                    ylabel={Accuracy},
                    symbolic x coords = {10,15,20,25,30,35,40,45,50},
                    xticklabel style={anchor= east,rotate=45 },
                    xtick=data,
                    ymax=80,
                    ymin=0,
                    legend pos=north west,
                    ymajorgrids=true,
                    grid style=dashed,
                    legend style={nodes={scale=1, transform shape}},
                    label style={font=\Large},
                    tick label style={font=\Large}
                ]
                \addplot+[mark size=3pt]
                    coordinates {
(10, 44.53125)
(15, 42.96875)
(20, 47.65625)
(25, 43.75)
(30, 48.4375)
(35, 48)
(40, 49)
(45, 49)
(50, 49)
                    };
                \addplot+[mark size=3pt]
                    coordinates {
(10, 50.78125)
(15, 53.90625)
(20, 43.75)
(25, 50.0)
(30, 51.5625)
(35, 50)
(40, 51.2)
(45, 51.4)
(50, 51.8)
                    };
                \addplot+[mark size=3pt]
                    coordinates {
(10, 53.125)
(15, 51.5625)
(20, 53.90625)
(25, 52.34375)
(30, 53.125)
(35, 52)
(40, 52)
(45, 53)
(50, 53)
                    };
                \addplot+[mark size=3pt]
                    coordinates {
(10, 46.09375)
(15, 44.53125)
(20, 47.65625)
(25, 47.65625)
(30, 48.75)
(35, 47)
(40, 46)
(45, 48)
(50, 49)
                    };
                \addplot+[mark size=3pt]
                    coordinates {
(10, 43.75)
(15, 45.3125)
(20, 48.4375)
(25, 48.4375)
(30, 47.3125)
(35, 48)
(40, 48)
(45, 49)
(50, 48)
                    };
                \addplot+[mark size=3pt]
                    coordinates {
(10, 32.1875)
(15, 32.96875)
(20, 32.96875)
(25, 35.1875)
(30, 34.1875)
(35, 34.2)
(40, 34.3)
(45, 35.3)
(50, 35.8)
                    };
                \addplot+[mark size=3pt]
                    coordinates {
(10, 31.9375)
(15, 32.0625)
(20, 33.28125)
(25, 33.84375)
(30, 34.28125)
(35, 34)
(40, 33)
(45, 34.2)
(50, 35.8)
                    };
                \addplot+[mark size=3pt]
                    coordinates {
(10, 32.8125)
(15, 32.15625)
(20, 30.9375)
(25, 33.59375)
(30, 34.03125)
(35, 34)
(40, 34)
(45, 35.8)
(50, 35.5)
                    };
                \addplot+[mark size=3pt]
                    coordinates {
(10, 31.71875)
(15, 31.71875)
(20, 32.5)
(25, 31.71875)
(30, 34.71875)
(35, 33)
(40, 32)
(45, 34.3)
(50, 34.8)
                    };
                \addplot+[mark size=3pt]
                    coordinates {
(10, 31.25)
(15, 31.25)
(20, 32.03125)
(25, 32.03125)
(30, 34.03125)
(35, 33)
(40, 33)
(45, 34)
(50, 34)
                    };
                \addplot+[mark size=3pt, color = green]
                    coordinates {
(10, 45)
(15, 46)
(20, 47)
(25, 48)
(30, 50)
(35, 50.2)
(40, 50.3)
(45, 50.8)
(50, 50.9)
                    };                     
                \end{axis}
            \end{tikzpicture}  
    }
    \caption{\tiny{OCTMNIST - 50\% malicious Clients}}
    \label{exp3a}
\end{subfigure}
~
~
~
\begin{subfigure}[tbh!]{0.45\columnwidth}
    \resizebox{1\columnwidth}{!}
    {
        \begin{tikzpicture}
                    \begin{axis}[
                    xlabel={Training Epoch},
                    ylabel={Accuracy},
                    symbolic x coords = {10,15,20,25,30,35,40,45,50},
                    %symbolic x coords = {1,5,10,15,20,25,30},
                    xticklabel style={anchor= east,rotate=45 },
                    xtick=data,
                    ymax=80,
                    ymin=0,
                    legend pos=north west,
                    ymajorgrids=true,
                    grid style=dashed,
                    legend style={nodes={scale=1, transform shape}},
                    label style={font=\Large},
                    tick label style={font=\Large}
                ]
                \addplot+[mark size=3pt]
                    coordinates {
(10, 54.91976351351351)
(15, 56.46114864864865)
(20, 56.84121621621622)
(25, 58.50929054054054)
(30, 58.34037162162162)
(35, 58.5)
(40, 59)
(45, 58.4)
(50, 58.2)
                    };\label{5c1}
                \addplot+[mark size=3pt]
                    coordinates {
(10, 54.7508445945946)
(15, 57.13682432432432)
(20, 57.72804054054054)
(25, 58.06587837837838)
(30, 59.332770270270274)
(35, 58.2)
(40, 58.9)
(45, 58.5)
(50, 59)
                    };\label{5c2}
                \addplot+[mark size=3pt]
                    coordinates {
(10, 54.43412162162162)
(15, 55.616554054054056)
(20, 57.326858108108105)
(25, 57.326858108108105)
(30, 58.27702702702703)
(35, 57)
(40, 59)
(45, 58)
(50, 58)
                    };\label{5c3}
                \addplot+[mark size=3pt]
                    coordinates {
(10, 55.29983108108109)
(15, 56.1866554054054)
(20, 57.64358108108109)
(25, 58.551520270270274)
(30, 58.973817567567565)
(35, 57)
(40, 58)
(45, 59)
(50, 59)
                    };\label{5c4}
                \addplot+[mark size=3pt]
                    coordinates {
(10, 55.616554054054056)
(15, 56.883445945945944)
(20, 58.12922297297297)
(25, 58.82601351351351)
(30, 58.720439189189186)
(35, 57)
(40, 58.4)
(45, 58.2)
(50, 58.1)
                    };\label{5c5}
                \addplot+[mark size=3pt]
                    coordinates {
(10, 29.56081081081081)
(15, 33.34037162162162)
(20, 33.973817567567565)
(25, 34.88175675675676)
(30, 34.670608108108105)
(35, 34)
(40, 34.342)
(45, 34.565)
(50, 34.23)
                    };\label{5c6}
                \addplot+[mark size=3pt]
                    coordinates {
(10, 31.967905405405407)
(15, 32.01013513513514)
(20, 32.8125)
(25, 34.98733108108108)
(30, 34.90287162162162)
(35, 33.12)
(40, 34.43)
(45, 35.546)
(50, 34.343)
                    };\label{5c7}
                \addplot+[mark size=3pt]
                    coordinates {
(10, 30.257601351351347)
(15, 32.157939189189186)
(20, 33.1714527027027)
(25, 34.83952702702703)
(30, 35.05067567567568)
(35, 34.7445)
(40, 34.34)
(45, 35.37)
(50, 34.45)
                    };\label{5c8}
                \addplot+[mark size=3pt]
                    coordinates {
(10, 31.12331081081081)
(15, 33.84712837837838)
(20, 34.33277027027027)
(25, 35.89527027027027)
(30, 35.97972972972973)
(35, 35)
(40, 34)
(45, 34)
(50, 35)
                    };\label{5c9}
                \addplot+[mark size=3pt]
                    coordinates {
(10, 31.038851351351347)
(15, 33.55152027027027)
(20, 33.973817567567565)
(25, 35.557432432432435)
(30, 35.282939189189186)
(35, 34)
(40, 34)
(45, 35)
(50, 35)
                    };\label{5c10}
                \addplot+[mark size=3pt, color = green]
                    coordinates {
(10, 36.14259133964817)
(15, 37.36679634641407)
(20, 38.72843369418133)
(25, 39.26336265223275)
(30, 39.43462449255751)
(35, 39.6)
(40, 39.8)
(45, 39.9)
(50, 40)
                    };\label{5gm}
                \end{axis}
            \end{tikzpicture}
    }
    \caption{\tiny{TissueMNIST - 50\% malicious Clients}}
    \label{exp3b}
\end{subfigure}
\caption{\footnotesize{Poisoning attack on FL architecture with 50\% malicious clients on OCTMNIST and TissueMNIST. The plot show the accuracies of benign 1 \ref{5c1}, benign 2 \ref{5c2}, benign 3 \ref{5c3}, benign 4 \ref{5c4}, benign 5 \ref{5c5}, malicious 6 \ref{5c6}, malicious 7 \ref{5c7}, malicious 8 \ref{5c8}, malicious 9 \ref{5c9}, malicious 10 \ref{5c10}, and, global evaluation \ref{5gm}.}}
\label{exp3}
\end{figure}
%%%%exp3

We then perform several experiments to show the effectiveness of the poisoning attack on a different percentage of malicious clients. To simulate a real-world scenario adversarial attack, we randomly selected a random client identified as malicious at the start of each experiment, and the other participants were identified as honest participants. In the experiment, we have set from 10\% to 50\% of the participants compromised to see the effect of the poisoned local model on the global model accuracy. We perform the label-flipping attack by choosing a specific class to poison by inverting its own label to another faulty label.

In Fig. \ref{exp2}, we perform a poisoning attack on our FL architecture that affected 10\% of the total clients. We apply the poisoning attack to both OCTMNIST and TissueMNIST. Under the same set-up, Fig. \ref{exp3} shows 50\% of the clients are malicious. From Fig. \ref{exp2a} and Fig. \ref{exp3a}, the average poisoned clients' accuracy is dropped to an average of 30\%. In OCTMNIST datasets, the poisoning attack can be reduced by up to 25\% on the client's accuracy. In the 10\% malicious client set-up, the global evaluation accuracy is dropped by 7\%. A 50\% malicious set-up can reduce the global model accuracy up to 22\%. 

In Fig. \ref{exp2b} and \ref{exp3b} show the average of adversarial clients accuracy is 32\%. For TissueMNIST, global model accuracy can reduce up to 9\% and 26\% for 10\% and 50\% malicious clients, respectively. However, the global model accuracy dropped significantly from 65\% to as low as 30\%. From Fig. \ref{exp2}, and Fig. \ref{exp3}, the poisoning attack we set up effectively decreases the client's accuracy and affects the global model.

%%%%exp4
\begin{figure}[tbh!]
\centering
\begin{subfigure}[tbh!]{0.45\columnwidth}
    \resizebox{1\columnwidth}{!}
    {
    \begin{tikzpicture}
                    \begin{axis}[
                    legend columns = 2
                    xlabel={Training Epoch},
                    ylabel={Accuracy},
                    symbolic x coords = {10,15,20,25,30,35,40,45,50},
                    xticklabel style={anchor= east,rotate=45 },
                    xtick=data,
                    ymax=80,
                    ymin=0,
                    legend pos=south east,
                    ymajorgrids=true,
                    grid style=dashed,
                    legend style={nodes={scale=1, transform shape}},
                    label style={font=\Large},
                    tick label style={font=\Large}
                ]
                \addplot+[mark size=3pt]
                    coordinates {
(10, 45.3125)
(15, 50.0)
(20, 53.90625)
(25, 50.78125)
(30, 56.25)
(35, 56.64)
(40, 57)
(45, 57.4)
(50, 57.8)
                    };
                \addplot+[mark size=3pt]
                    coordinates {
(10, 53.125)
(15, 53.90625)
(20, 46.875)
(25, 56.25)
(30, 57.03125)
(35, 56.21)
(40, 56.8)
(45, 57.2)
(50, 57.8)
                    };
                \addplot+[mark size=3pt]
                    coordinates {
(10, 46.875)
(15, 50.0)
(20, 50.0)
(25, 51.5625)
(30, 50.78125)
(35, 50.8)
(40, 51.2)
(45, 51.6)
(50, 51.9)
                    };
                \addplot+[mark size=3pt]
                    coordinates {
(10, 46.09375)
(15, 46.875)
(20, 44.53125)
(25, 50.0)
(30, 50.78125)
(35, 50.9)
(40, 51.2)
(45, 51.8)
(50, 52.1)
                    };
                \addplot+[mark size=3pt]
                    coordinates {
(10, 52.34375)
(15, 52.34375)
(20, 52.34375)
(25, 53.90625)
(30, 53.125)
(35, 52.2)
(40, 52.4)
(45, 53.2)
(50, 53.4)
                    };
                \addplot+[mark size=3pt]
                    coordinates {
(10, 50.78125)
(15, 50.0)
(20, 50.0)
(25, 51.5625)
(30, 50.78125)
(35, 50.3)
(40, 51.2)
(45, 51.7)
(50, 52)
                    };
                \addplot+[mark size=3pt]
                    coordinates {
(10, 43.75)
(15, 43.75)
(20, 47.65625)
(25, 43.75)
(30, 52.34375)
(35, 51.8)
(40, 51.9)
(45, 52.2)
(50, 52.3)
                    };
                \addplot+[mark size=3pt]
                    coordinates {
(10, 46.875)
(15, 47.65625)
(20, 50.0)
(25, 50.78125)
(30, 51.5625)
(35, 51.23)
(40, 51.86)
(45, 52.223)
(50, 52.785)

                    };
                \addplot+[mark size=3pt]
                    coordinates {
(10, 46.09375)
(15, 51.5625)
(20, 51.5625)
(25, 57.03125)
(30, 55.46875)
(35, 55)
(40, 56)
(45, 57.5)
(50, 57.8)
                    };
                \addplot+[mark size=3pt]
                    coordinates {
                    (10, 0)

                    };
                \addplot+[mark size=3pt, color = green]
                    coordinates {
(10, 62)
(15, 66)
(20, 67)
(25, 69)
(30, 72)
(35, 72.2)
(40, 72.5)
(45, 72.8)
(50, 73)
                    }; 
                \legend{Benign 1,Benign 2,Benign 3,Benign 4,Benign 5,Benign 6,Benign 7,Benign 8,Benign 9,,Global Evaluation}
                \end{axis}
            \end{tikzpicture} 
    }
    \caption{\tiny{OCTMNIST - Filtered 10\% malicious Clients}}
    \label{exp4a}
\end{subfigure}
~
~
~
\begin{subfigure}[tbh!]{0.45\columnwidth}
    \resizebox{\columnwidth}{!}
    {
    \begin{tikzpicture}
                    \begin{axis}[
                    legend columns =2
                    xlabel={Training Epoch},
                    ylabel={Accuracy},
                    symbolic x coords = {10,15,20,25,30,35,40,45,50},
                    xticklabel style={anchor= east,rotate=45 },
                    xtick=data,
                    ymax=80,
                    ymin=0,
                    legend pos=south east,
                    ymajorgrids=true,
                    grid style=dashed,
                    legend style={nodes={scale=1, transform shape}},
                    label style={font=\Large},
                    tick label style={font=\Large}
                ]
                \addplot+[mark size=3pt]
                    coordinates {
(10, 44.53125)
(15, 42.96875)
(20, 47.65625)
(25, 43.75)
(30, 48.4375)
(35, 49.6)
(40, 50.2)
(45, 51.3)
(50, 51.8)
                    };
                \addplot+[mark size=3pt]
                    coordinates {
(10, 50.78125)
(15, 53.90625)
(20, 43.75)
(25, 50.0)
(30, 51.5625)
(35, 51.8)
(40, 51.9)
(45, 52.1)
(50, 52.2)
                    };
                \addplot+[mark size=3pt]
                    coordinates {
(10, 53.125)
(15, 51.5625)
(20, 53.90625)
(25, 52.34375)
(30, 53.125)
(35, 53.2)
(40, 52.9)
(45, 53.2)
(50, 53.5)
                    };
                \addplot+[mark size=3pt]
                    coordinates {
(10, 46.09375)
(15, 44.53125)
(20, 47.65625)
(25, 47.65625)
(30, 47.75)
(35, 49.23)
(40, 50.2)
(45, 51.212)
(50, 51.972)
                    };
                \addplot+[mark size=3pt]
                    coordinates {
(10, 43.75)
(15, 45.3125)
(20, 48.4375)
(25, 48.4375)
(30, 45.3125)
(35, 47.98)
(40, 59.73)
(45, 50.1)
(50, 50.689)
                    };
                \addplot+[mark size=3pt]
                    coordinates {
(10,0)
                    };
                \addplot+[mark size=3pt]
                    coordinates {
(10,0)
                    };
                \addplot+[mark size=3pt]
                    coordinates {
(10,0)
                    };
                \addplot+[mark size=3pt]
                    coordinates {
(10,0)
                    };
                \addplot+[mark size=3pt]
                    coordinates {
(10,0)
                    };
                \addplot+[mark size=3pt, color = green]
                    coordinates {
                    (10, 59.23)
                    (15, 60.453)
                    (20, 65.344)
                    (25, 70.33)
                    (30, 70.47)
                    (35, 70.8343)
                    (40, 70.974543)
                    (45, 71.863)
                    (50, 71.972)
                    };
                    \legend{Benign 1,Benign 2,Benign 3,Benign 4,Benign 5,,,,,,Global Evaluation}
                \end{axis}
            \end{tikzpicture}
    }
    \caption{\tiny{OCTMNIST - Filtered 50\% malicious Clients}}
    \label{exp4b}
\end{subfigure}
~
~
~
\begin{subfigure}[tbh!]{0.45\columnwidth}
    \resizebox{1\columnwidth}{!}
    {
    \begin{tikzpicture}
                    \begin{axis}[
                    legend columns=2
                    xlabel={Training Epoch},
                    ylabel={Accuracy},
                    symbolic x coords = {10,15,20,25,30,35,40,45,50},
                    %symbolic x coords = {1,5,10,15,20,25,30},
                    xticklabel style={anchor= east,rotate=45 },
                    xtick=data,
                    ymax=80,
                    ymin=0,
                    legend pos=south east,
                    ymajorgrids=true,
                    grid style=dashed,
                    legend style={nodes={scale=1, transform shape}},
                    label style={font=\Large},
                    tick label style={font=\Large}
                ]
                \addplot+[mark size=3pt]
                    coordinates {
(10, 56.9679054054054)
(15, 59.35388513513513)
(20, 59.079391891891895)
(25, 60.34628378378378)
(30, 59.90287162162162)
(35, 60.432)
(40, 60.332)
(45, 60.23)
(50, 59.952)
                    };
                \addplot+[mark size=3pt]
                    coordinates {
(10, 55.13091216216216)
(15, 58.50929054054054)
(20, 59.016047297297305)
(25, 59.94510135135135)
(30, 59.88175675675676)
(35, 60.12)
(40, 59.55)
(45, 59.776)
(50, 60.1)
                    };
                \addplot+[mark size=3pt]
                    coordinates {
(10, 55.173141891891895)
(15, 57.60135135135135)
(20, 57.64358108108109)
(25, 59.501689189189186)
(30, 59.26942567567568)
(35, 59.231)
(40, 60.21)
(45, 60.428)
(50, 60.9877)
                    };
                \addplot+[mark size=3pt]
                    coordinates {
(10, 56.376689189189186)
(15, 60.17736486486487)
(20, 59.860641891891895)
(25, 60.8741554054054)
(30, 60.66300675675676)
(35, 60.8973)
(40, 61.12)
(45, 60.872)
(50, 61.762)
                    };
                \addplot+[mark size=3pt]
                    coordinates {
(10, 55.65878378378378)
(15, 58.467060810810814)
(20, 58.06587837837838)
(25, 60.0929054054054)
(30, 59.501689189189186)
(35, 60.422)
(40, 60.322)
(45, 60.765)
(50, 61.1)
                    };
                \addplot+[mark size=3pt]
                    coordinates {
(10, 57.05236486486487)
(15, 59.45945945945946)
(20, 60.70523648648649)
(25, 61.61317567567568)
(30, 61.570945945945944)
(35, 60.7623)
(40, 61.232)
(45, 60.978)
(50, 61.9762)
                    };
                \addplot+[mark size=3pt]
                    coordinates {
(10, 56.08108108108109)
(15, 57.8758445945946)
(20, 59.64949324324324)
(25, 59.755067567567565)
(30, 60.17736486486487)
(35, 60.12112)
(40, 60.734)
(45, 59.76)
(50, 60.25)
                    };
                \addplot+[mark size=3pt]
                    coordinates {
(10, 55.4054054054054)
(15, 59.90287162162162)
(20, 59.776182432432435)
(25, 60.641891891891895)
(30, 60.43074324324324)
(35, 59.21)
(40, 59.98)
(45, 60.1)
(50, 60.124)
                    };
                \addplot+[mark size=3pt]
                    coordinates {
(10, 56.735641891891895)
(15, 58.445945945945944)
(20, 59.079391891891895)
(25, 59.64949324324324)
(30, 59.248310810810814)
(35, 58.345)
(40, 59.645)
(45, 58.213)
(50, 59.75)
                    };
                \addplot+[mark size=3pt]
                    coordinates {
(10,0)
                    };
                \addplot+[mark size=3pt, color = green]
                    coordinates {
                    (10, 60.6)
                    (15, 61.9)
                    (20, 62.4)
                    (25, 62.9)
                    (30, 63.0)
                    (35, 63.122)
                    (40, 63.23)
                    (45, 63.264)
                    (50, 63.487)
                    };
                    \legend{Benign 1,Benign 2,Benign 3,Benign 4,Benign 5,Benign 6,Benign 7,Benign 8,Benign 9,,Global Evaluation}
                \end{axis}
            \end{tikzpicture}
    }
    \caption{\tiny{TissueMNIST - Filtered 10\% malicious Clients}}
    \label{exp4c}
\end{subfigure}
~
~
~
\begin{subfigure}[tbh!]{0.45\columnwidth}
    \resizebox{\columnwidth}{!}
    {
    \begin{tikzpicture}
                    \begin{axis}[
                    legend columns=2
                    xlabel={Training Epoch},
                    ylabel={Accuracy},
                    symbolic x coords = {10,15,20,25,30,35,40,45,50},
                    xticklabel style={anchor= east,rotate=45 },
                    xtick=data,
                    ymax=80,
                    ymin=0,
                    legend pos=south east,
                    ymajorgrids=true,
                    grid style=dashed,
                    legend style={nodes={scale=1, transform shape}},
                    label style={font=\Large},
                    tick label style={font=\Large}
                ]
                \addplot+[mark size=3pt]
                    coordinates {
(10, 54.91976351351351)
(15, 56.46114864864865)
(20, 56.84121621621622)
(25, 58.50929054054054)
(30, 58.34037162162162)
(35, 58.746)
(40, 59.32)
(45, 59.87)
(50, 59.98)
                    };
                \addplot+[mark size=3pt]
                    coordinates {
(10, 54.7508445945946)
(15, 57.13682432432432)
(20, 57.72804054054054)
(25, 58.06587837837838)
(30, 59.332770270270274)
(35, 59.862)
(40, 59.98)
(45, 60.212)
(50, 60.332)

                    };
                \addplot+[mark size=3pt]
                    coordinates {
(10, 54.43412162162162)
(15, 55.616554054054056)
(20, 57.326858108108105)
(25, 57.326858108108105)
(30, 58.27702702702703)
(35, 58.552)
(40, 58.212)
(45, 58.622)
(50, 58.87)
                    };
                \addplot+[mark size=3pt]
                    coordinates {
(10, 55.29983108108109)
(15, 56.1866554054054)
(20, 57.64358108108109)
(25, 58.551520270270274)
(30, 58.973817567567565)
(35, 59.2)
(40, 59.421)
(45, 59.877)
(50, 60.2)
                    };
                \addplot+[mark size=3pt]
                    coordinates {
(10, 55.616554054054056)
(15, 56.883445945945944)
(20, 58.12922297297297)
(25, 58.82601351351351)
(30, 58.720439189189186)
(35, 59.12)
(40, 59.433)
(45, 59.522)
(50, 59.651)
                    };
                \addplot+[mark size=3pt]
                    coordinates {
(10,0)
                    };
                \addplot+[mark size=3pt]
                    coordinates {
(10,0)
                    };
                \addplot+[mark size=3pt]
                    coordinates {
(10,0)
                    };
                \addplot+[mark size=3pt]
                    coordinates {
(10,0)
                    };
                \addplot+[mark size=3pt]
                    coordinates {
(10,0)
                    };
                \addplot+[mark size=3pt, color = green]
                    coordinates {
                    (10, 59.45)
                    (15, 60.92)
                    (20, 61.475)
                    (25, 62.21)
                    (30, 62.01)
                    (35, 62.35)
                    (40, 62.52)
                    (45, 62.86)
                    (50, 63.2)

                    };
                    \legend{Benign 1,Benign 2,Benign 3,Benign 4,Benign 5,,,,,,Global Evaluation}
                \end{axis}
            \end{tikzpicture}
    }
    \caption{\tiny{TissueMNIST - Filtered 50\% malicious Clients}}
    \label{exp4d}
\end{subfigure}
\caption{\footnotesize{Secure model verifier effectiveness on OCTMNIST and TissueMNIST for filtering various percentage of malicious clients}}
\label{exp4}
\end{figure}
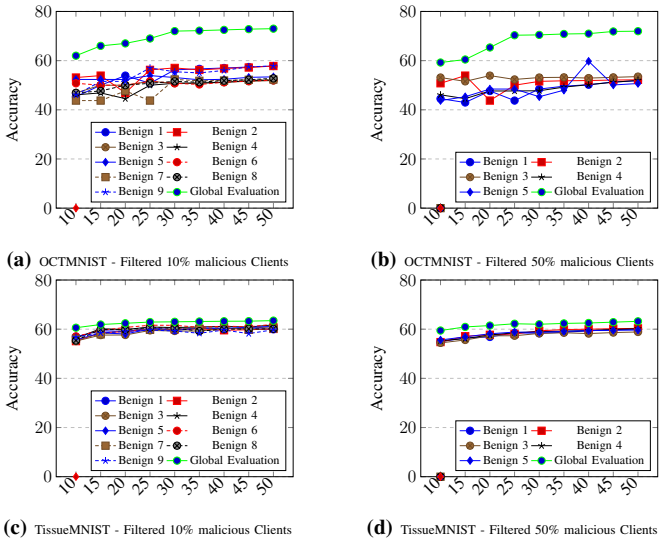
%%%%exp4

In Fig. \ref{exp4}, we show the effectiveness of our proposed defense method. In this experiment, we perform a secure model verifier to remove the poisoning attack from various amounts of malicious clients before the aggregation process. At first, we removed the 10\% malicious clients from both datasets. Based on Fig \ref{exp4a} and Fig \ref{exp4c}, the global model can have the same accuracy as shown in Fig \ref{exp1}. In the 50\% of malicious client set-up, OCTMNIST and TissueMNIST can recover up to 25\% accuracy for the global model. From this experiment, our secure model verifier can detect from a low percentage to half of the malicious clients and exclude them from the aggregation process. The result of our filtration process can improve the global model evaluation accuracy back to the normal state.

%%%%exp5
\begin{figure}[tbh!]
\centering
\begin{subfigure}[tbh!]{0.4\columnwidth}
    \resizebox{1\columnwidth}{!}{
            \begin{tikzpicture}
                \begin{axis}[
                    xlabel={Number of Images},
                    ylabel={Time(s)},
                    symbolic x coords = {5,10,15,20,30},
                    xticklabel style={anchor= east,rotate=45 },
                    xtick=data,
                    ymax=2000,
                    ymin=0,
                    legend pos=north west,
                    ymajorgrids=true,
                    grid style=dashed,
                    legend style={nodes={scale=1, transform shape}},
                    label style={font=\Large},
                    tick label style={font=\Large}
                ]
                \addplot+[mark size=3pt]
                    coordinates {
                    (5,215)
                    (10,522)
                    (15,792)
                    (20,1011)
                    (30,1278)
                    };
                \addplot+[mark size=3pt]
                    coordinates {
                    (5,355)
                    (10,620)
                    (15,954)
                    (20,1351)
                    (30,1602)
                    };
                \legend{Unencrypted, Encrypted}
                \end{axis}
            \end{tikzpicture}
            }
    \caption{\tiny{OCTMNIST}}
    \label{exp5a}
\end{subfigure}
~
~
~
\begin{subfigure}[tbh!]{0.4\columnwidth}
    \resizebox{1\columnwidth}{!}{
        \begin{tikzpicture}
                \begin{axis}[
                    xlabel={Number of Images},
                    ylabel={Time(s)},
                    symbolic x coords = {5,10,15,20,30},
                    xticklabel style={anchor= east,rotate=45 },
                    xtick=data,
                    ymax=2000,
                    ymin=0,
                    legend pos=north west,
                    ymajorgrids=true,
                    grid style=dashed,
                    legend style={nodes={scale=1, transform shape}},
                    label style={font=\LARGE},
                    tick label style={font=\Large}
                ]
                \addplot+[mark size=3pt]
                    coordinates {
                    (5,220)
                    (10,523)
                    (15,815)
                    (20,988)
                    (30,1311)
                    };
                \addplot+[mark size=3pt]
                    coordinates {
                    (5,372)
                    (10,645)
                    (15,986)
                    (20,1298)
                    (30,1552)
                    };
                \legend{Unencrypted, Encrypted}
                \end{axis}
            \end{tikzpicture}
        }
    \caption{\tiny{TissueMNIST}}
    \label{exp5b}
\end{subfigure}
\caption{\footnotesize{Time cost comparison for normal inference and encrypted inference process with a)OCTMNIST; b)TissueMNIST;}}
\label{exp5}
\end{figure}
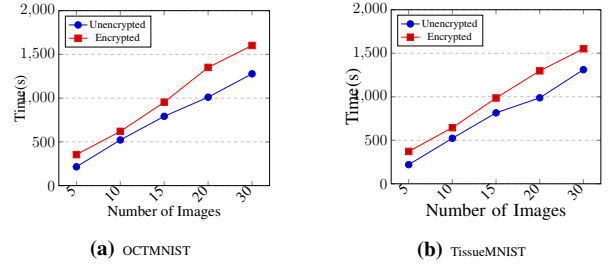
%%%%exp5

Our secure model verifier method uses the encrypted inference method to provide local model privacy. In Fig. \ref{exp5}, we compare the time cost for the inference process between unencrypt and encrypted images. The result shows that OCTMNIST and TissueMNIST's inference process has approximately similar time costs. From our analysis, it's because we are using the same model, and the model holds the same number of parameters. In Fig. \ref{exp5a} and Fig. \ref{exp5b}, the time cost is increasing linearly, and the average time difference is 200 seconds for both unencrypt and encrypted images. From the given result, our method can preserve the privacy of the local model without affecting any significant impact on the performance.

%%%%exp6
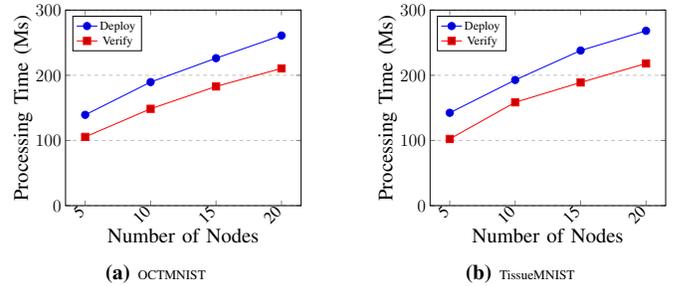
\begin{figure}[tbh!]
\centering
\begin{subfigure}[tbh!]{0.45\columnwidth}
    \resizebox{1\columnwidth}{!}{
        \begin{tikzpicture}
                \begin{axis}[
                    xlabel={Number of Nodes},
                    ylabel={Processing Time (Ms)},
                    symbolic x coords = {5,10,15,20},
                    xticklabel style={anchor= east,rotate=45 },
                    xtick=data,
                    ymax=300,
                    ymin=0,
                    legend pos=north west,
                    ymajorgrids=true,
                    grid style=dashed,
                    legend style={nodes={scale=1, transform shape}},
                    label style={font=\LARGE},
                    tick label style={font=\Large}
                ]
                \addplot+[mark size=3pt]
                    coordinates {
                        (5,139.31)
                        (10,189.543)
                        (15,226.231)
                        (20,260.983)
                    };
                \addplot+[mark size=3pt]
                    coordinates {
                        (5,105.432)
                        (10,148.543)
                        (15,182.869)
                        (20,210.544)
                    };
                \legend{Deploy, Verify}
                \end{axis}
            \end{tikzpicture}
        }
    \caption{\tiny{OCTMNIST}}
    \label{exp6a}
\end{subfigure}
~
~
~
\begin{subfigure}[tbh!]{0.45\columnwidth}
    \resizebox{1\columnwidth}{!}{
        \begin{tikzpicture}
                \begin{axis}[
                    xlabel={Number of Nodes},
                    ylabel={Processing Time (Ms)},
                    symbolic x coords = {5,10,15,20},
                    xticklabel style={anchor= east,rotate=45 },
                    xtick=data,
                    ymax=300,
                    ymin=0,
                    legend pos=north west,
                    ymajorgrids=true,
                    grid style=dashed,
                    legend style={nodes={scale=1, transform shape}},
                    label style={font=\LARGE},
                    tick label style={font=\Large}
                ]
                \addplot+[mark size=3pt]
                    coordinates {
                        (5,142.421)
                        (10,192.782)
                        (15,237.98)
                        (20,268.233)
                    };
                \addplot+[mark size=3pt]
                    coordinates {
                        (5,102.324)
                        (10,158.522)
                        (15,188.987)
                        (20,218.112)
                    };
                \legend{Deploy, Verify}
                \end{axis}
            \end{tikzpicture}
        }
    \caption{\tiny{TissueMNIST}}
    \label{exp6b}
\end{subfigure}
\caption{\footnotesize{Processing time for adding the global model to the blockchain with different number of blockchain nodes.}}
\label{exp6}
\end{figure}
%%%%exp6

In this experiment, the blockchain node aggregates the models from the secure verifier to generate the global model. Then the verified model is deployed to the blockchain network. Fig. \ref{exp6} shows the time required to execute both the verification and deployment of the global model to the blockchain. Our experiment tested the performance using several blockchain nodes ranging from 5 to 20 nodes. From the given result, the deployment phase is more comprehensive, requiring approximately 140ms to 270 ms across the 20 nodes. The verification phase is faster, starting from 100ms to 220 ms, and both processes show a marginal increase in the time taken as more nodes are added to the blockchain network.

\subsection{Discussion}
In this section, we summarize the performance of our proposed method. As discussed in Section \ref{sec:results}, we conducted a series of experiments to evaluate the efficacy of our proposed method. Based on the empirical results, the following conclusions can be drawn.

\begin{itemize}

    \item \textbf{Privacy of Local Dataset}:
    The federated learning scenario allows each participant to collaboratively train the machine learning model locally with their local datasets. Later the machine learning model will send to the cloud for the model aggregation process. The federated learning method is unlike the centralized machine learning approach, where the participants' local data needs to be sent to the cloud for the learning process. Therefore the federated learning scenario can ensure the privacy of the participant's sensitive datasets.
    
    \item \textbf{Robustness of Local and Global Model}:
    In our scenario, we consider an adversary that performs poisoning attacks on the participant's datasets. The poisoning attack will lead to a faulty local model and a poisoned global model. Early detection to eliminate any poisoned local model to be excluded from the aggregation process is required. From the experiment results, our proposed work can eliminate all the poisoned participant models. From this result, our architecture can guarantee the robustness of the participant's local and global models.

    \item \textbf{Privacy of Local Model}:
    In our framework, we perform an inference process in the cloud to verify whether or not the local model is compromised. However, an attacker can perform a membership inference attack \cite{8835245} on the participant's model and leak sensitive data from the model. Therefore, we leverage the SMPC-based encrypted inference process to protect the local model from the attacker while verifying the local model from poisoning attacks. As the local model is protected using SMPC-based secret share protocol, model inversion attacks \cite{fredrikson2015model}, and parameter stealing \cite{8418595} cannot be performed on a local model by an attacker.
    
    \item \textbf{SMPC-based Secure Aggregation}:
    Participants' local models are collected and aggregated in the global model in federated learning. The aggregation process is the core step of federated learning to achieve a higher accuracy learning model. However, aggregation is typically performed on a regular server. Previously, \cite{wei2020federated,9253545} proposed the differential privacy (DP) method to secure the model from membership inference attack \cite{8835245}. Nevertheless, DP will significantly drop the accuracy of the global model. As the aggregation is performed in the blockchain node using SMPC-based secure aggregation, the adversaries cannot tamper with the aggregation process while maintaining the model's accuracy.
    
    \item \textbf{Verifiablity of the Global Model}:
    Blockchain is a well-known decentralized technology that can maintain data integrity. The data must be verified among the blockchain node using a consensus mechanism to store data in the blockchain. Once the data is verified, the blockchain will create a new block to store the data. Since all blockchain nodes stored have the same ledger, adversaries cannot tamper with the integrity of the data. In our proposed framework, we leverage blockchain to store the latest global model after the secure aggregation process. The decentralized process makes it impossible for adversaries to tamper with or alter the global model since it will change the hash value. Later, the global model stored in the blockchain will be sent to the federated learning participants. Moreover, the participants can verify the integrity of the global model by checking the signatures and hashes before they use it for the inference process.

\end{itemize}

\section{Future Research direction}\label{sec:fut}
This paper introduces Blockchain-based federated learning with SMPC model verification to overcome an adversarial attack. Hence, there are several challenges requiring further research:

\textbf{Efficient consensus mechanism:} 
The use of blockchain can ensure the integrity of the data from the malicious attacker. However, the consensus mechanism and the synchronization consume substantial computational power. Therefore, developing an efficient consensus mechanism to reduce computational and energy resources is a topic that needs to be explored.

\textbf{Expensive communication:}
Secure multi-party computation can guarantee the privacy and security of multiple participants. However, the SMPC needs to be performed on several communication rounds and involves multiple parties while performing computation. Federated learning also requires multiple rounds to achieve the best accuracy for the global model. A communication-efficient mechanism must be developed to make the FL and SMPC more efficient in a practical scenario.

\section{Conclusion}\label{sec:con}
This paper proposes blockchain-based federated learning with a secure model verification for securing healthcare systems. The main objective is to ensure the local model is poisoned-free while maintaining privacy and providing verifiability for the federated learning participants.

In this framework, we perform a privacy-preserving verification process on the local model before the aggregation process. To preserve privacy on the local model, the verification is performed through an encrypted inference supported by SMPC protocol. This method allows the verifier to check the model with encrypted models and images. Once the local model is verified, the verified share of the local model is sent to the blockchain node. Blockchain and the hospital will perform SMPC-based secure aggregation. Once the majority of nodes have the same result, the global model is stored in the blockchain. Later, the tamper-proof storage will distribute the updated global model to every hospital that joins the federated learning round.

In the experiment, we use Convolutional Neural Network (CNN) based algorithms with several medical datasets to generate local models and aggregate them under FL settings. Our experiment results show that the model encrypted verification process can eliminate all the participants' poisoned models while maintaining the privacy of the local model. In addition, we can recover up to 25\% for the global model accuracy. It is essential to mention that our secure inference processing time is almost similar to the original inference process.

In the future, we plan to develop an efficient consensus mechanism for blockchain-based aggregation. In this paper, we assume that all hospitals use the homogeneous model and use the same setup to generate their respective local models. However, we plan to broaden our work in the future to support a heterogeneous model in blockchain-based federated learning.

\section{ACKNOWLEDGMENTS}
This work is supported by the Australian Research Council Discovery Project (DP210102761).

% \ifCLASSOPTIONcaptionsoff
%   \newpage
% \fi
% \newpage
\bibliographystyle{IEEEtran}
\bibliography{References}

\end{document}